\documentclass[10pt,times,a4paper,floatfix]{iopart}
\usepackage{epsfig}
\usepackage{graphics}
\usepackage{graphicx}
\usepackage{bm}
\usepackage{iopams}
\usepackage[dvipsnames]{xcolor}
\usepackage{bm}
\usepackage{times}
\usepackage{xspace}
\usepackage[varg]{txfonts}
\usepackage[normalem]{ulem} 
\usepackage[colorlinks]{hyperref}
\usepackage[caption=false]{subfig}
\usepackage{booktabs}
\usepackage{tabularx}
\usepackage[utf8]{inputenc}

\definecolor{LinkColor}{rgb}{0.75, 0, 0}
\definecolor{CiteColor}{rgb}{0, 0.5, 0.5}
\definecolor{UrlColor}{rgb}{0, 0, 0.75}
\hypersetup{linkcolor=LinkColor}
\hypersetup{citecolor=CiteColor}
\hypersetup{urlcolor=UrlColor}
\usepackage{perpage}
\MakePerPage{footnote}

\newcommand{\eqref}[1]{\ref{eqn: #1}}

\newcommand{\be}{\begin{equation}}
\newcommand{\ee}{\end{equation}}

\begin{document}

\title[Quantum black hole spectroscopy]{Quantum black hole spectroscopy: probing the quantum nature of the black hole area using LIGO-Virgo ringdown detections}

\author{Danny Laghi$^{1,2}$, Gregorio Carullo$^{1,2}$, John Veitch$^{3}$, Walter Del Pozzo$^{1,2}$}
\address{${}^1$ Dipartimento di Fisica ``Enrico Fermi'', Universit\`a di Pisa, Pisa I-56127, Italy}
\address{${}^2$ INFN sezione di Pisa, Pisa I-56127, Italy}
\address{${}^3$ Institute for Gravitational Research, University of Glasgow, Glasgow, G12 8QQ, United Kingdom}

\ead{danny.laghi@df.unipi.it}

\date{\today}

\begin{abstract}
We present a thorough observational investigation of the heuristic quantised ringdown model presented in \cite{FOIT-KLEBAN}. This model is based on the Bekenstein-Mukhanov conjecture, stating that the area of a black hole horizon is an integer multiple of the Planck area~$l_P^2$ multiplied by a phenomenological constant,~$\alpha$, which can be viewed as an additional black hole intrinsic parameter. 
Our approach is based on a time-domain analysis of the gravitational wave signals produced by the ringdown phase of binary black hole mergers detected by the LIGO and Virgo collaboration. 
Employing a full Bayesian formalism and taking into account the complete correlation structure among the black hole parameters, we show that the value of $\alpha$ cannot be constrained using only GW150914, in contrast to what was suggested in~\cite{FOIT-KLEBAN}. We proceed to repeat the same analysis on the new gravitational wave events detected by the LIGO and Virgo Collaboration up to 1 October 2019, obtaining a combined-event measure equal to $\alpha = 15.6^{+20.5}_{-13.3}$ and a combined log odds ratio of $0.1 \pm 0.6$, implying that 
current data are not informative enough to favour or discard this model against general relativity.
We then show that using a population of $\mathcal{O}(20)$ GW150914-like simulated events -- detected by the current infrastructure of ground-based detectors at their design sensitivity -- it is possible to confidently falsify the quantised model or prove its validity, in which case probing $\alpha$ at the few \% level. Finally we classify the stealth biases that may show up in a population study.
\end{abstract}

\maketitle


\section{Introduction}

Black holes (BHs) are amongst the simplest and yet most mysterious and fascinating objects of our Universe. 
Until the direct observation of gravitational waves (GW), they could only be explored indirectly, through their influence on their surroundings~\cite{Bambi:2020cyv, ABUTER2020}. We recently had the opportunity to look at the exterior region surrounding a black hole~\cite{EHT-2019-1}.
The observation of GW150914~\cite{GW150914} by the two LIGO instruments~\cite{LIGO} and the following signals from LIGO and Virgo~\cite{VIRGO}, 
provided a new exciting avenue to gain unprecedented insights on the space-time dynamics and test key ideas in gravitational physics \cite{WILL2018}.

GWs emitted by BHs interact feebly with matter and reach the detectors on Earth unimpeded, providing a natural laboratory to probe fundamental laws of physics \cite{BERTI_TGR_2015}. 
Their shape carries information about the physical properties of the source space-time, allowing measurement of the parameters characterising the BHs as well as constraints on potential departures from the predictions of general relativity (GR)~\cite{TGR-LVC2016, TGR-170817, TGR-O1-O2, O3a-TGR-PAPER}, paving the way for future tests on the existence of non-BH compact objects~\cite{LIEBLING-PALENZUELA, MAZUR-MOTTOLA}. 
It is tempting to ask whether GW physics could shed light on theoretical ideas about the quantum properties of BHs. 
Being vacuum, pure space-time solutions to GR, BHs play a very important role in the development of a theory of quantum gravity~\cite{BEKENSTEIN1973,  MUKHANOV, MAZUR1987, MAGGIORE1994, KOGAN1986, DANIELSSON-SCHIFFER1993, MAKELA1997, HOD, DAVIDSON} and, in principle, could help us to test quantum features of space-time, if any imprint is left on classical observables. 
Among the most promising candidates to detect these putative effects is the signal emitted during the latest stages of a binary BH (BBH) merger, the ``ringdown'' phase, where perturbation theory can be applied to study the Einstein field equations \cite{CHANDRA, KOKKOTAS-SCHMIDT, NOLLERT, TEUKOLSKY-1972, TEUKOLSKY-1973}. 
This phase has been shown to be well-suited for investigation on the nature of BHs~\cite{Detweiler:1980gk}, giving rise to a subfield of GW physics known as ``BH spectroscopy'' \cite{Dreyer:2003bv, LISA_spectroscopy, Gossan:2011ha, PhysRevD.90.064009, Carullo:2018sfu, Brito:2018rfr, Carullo:2019flw, Isi:2019aib, Bhagwat:2019bwv, Bhagwat:2019dtm}.
A ringdown signal is a space-time oscillation modelled as a linear combination of damped oscillations followed by late-time, hitherto undetected, power-law tails.
Each mode of these oscillations is called quasinormal mode (QNM).
In GR, the QNMs of an astrophysical Kerr BH~\cite{KERR} are uniquely determined by the mass and the angular momentum of the BH, according to the ``no hair'' conjecture~\cite{Penrose:2002col, 2002nmgm.meet...28K, Chrusciel:2012jk, Pretorius2017} stating that the physical spectrum of QNMs is exclusively determined by conserved quantities.
These quantities are subject to a Gauss' law, implying that they can be determined by measurements from afar.
A substantial amount of theoretical work has been done on the study of additional hairs, and results in four-dimensional space-time seem to corroborate the conjecture~\cite{ISRAEL, CARTER, BEKENSTEIN-NOHAIR-PRL, BEKENSTEIN-NOHAIR-PRD, ROBINSON}, although no formal proof exists yet. 

Possible new properties of BHs could naturally appear if one considers quantum corrections to the BH structure. 
A glimpse of how a quantum theory of gravity may radically change the classical view of BHs comes from the quantum mechanics of BH event horizons. 
Notable features are the Hawking radiation and the entropy-area relation, expected to be distinctive trademarks of a full quantum BH (QBH) theory.
In a tentative heuristic description of QBHs, one would expect that the fundamental quantities describing macroscopic classical BHs could play the role of unique quantum numbers characterizing their properties~\cite{BEKENSTEIN1996,BEKENSTEIN1997QUANTUM}. 
This route was undertaken by Bekenstein who, since the discovery of the entropy-area law, adopted a partial scheme to quantize BHs through the quantisation of the classical hairs of a BH: mass, electric charge, magnetic monopole, and angular momentum (which, in agreement with GW physics conventions, we will refer to as ``spin''), deriving their eigenstates for a non-extremal and stationary BH~\cite{BEKENSTEIN1973, BEKENSTEIN1974}. 
Bekenstein's scheme does not specify the microphysics responsible for the quantisation of these quantities; 
however, assuming standard commutation relations for this set of operators, the mass eigenvalues as well as the area spectrum can be derived from a quantum operator algebra. 
This result led to what has been later called Bekenstein-Mukhanov (BM) area conjecture \cite{BEKENSTEIN1974, MUKHANOV, BEKENSTEIN-MUKHANOV}, which relies on the observation that the horizon area behaves classically as an adiabatic invariant~\cite{CHRISTODOULOU-RUFFINI}: Ehrenfest's hypothesis~\cite{KLEIN1959} can then be invoked to state then the horizon area should be quantized in any theory of quantum gravity~\cite{BEKENSTEIN1997QUANTUM}. 
In classical GR, the mass spectrum of BHs is a continuum, meaning that any value of mass can be assumed by a BH.
The BM scheme suggests that, in a presumed quantum theory of BHs,
the mass spectrum must be discrete and degenerate; accordingly, a QBH can only absorb and emit quanta of any species at specific discrete frequencies. 
A notable fact is that the frequencies of the QBH depend explicitly on a phenomenological
parameter, that we will label $\alpha$. 
In their original formulation, nothing is said about which form the QNMs of a quantized BH should assume.
Foit and Kleban (FK henceforth) recently proposed a heuristic link between the BM theory and ringdown phenomenology~\cite{FOIT-KLEBAN}. The authors posit the BM area conjecture in the context of BBH GW physics: under specific assumptions (reviewed in Sec.~\ref{sec:BM conjecture_section}), one can think of the ringdown of a BM QBH as being affected by the BH quantized spectrum derived from the area conjecture. As we will show, this may lead to a change in the fundamental (i.e. longest-lived) QNM, hence making the conjecture testable against LIGO-Virgo data.

In this paper, we investigate whether such a proposal can indeed be tested with ringdown signals. 
We will not discuss or question the validity of the assumptions of the FK model, but instead we will use it as a paradigmatic example of QBH model to find out whether a test of this kind is possible \emph{in principle} with current GW detector facilities.
In our proof-of-concept model, the fundamental QNM observed during the ringdown is identified with the quantized fundamental frequency of the QBH. This allows us to constrain the quantum parameter, $\alpha$, with Bayesian parameter estimation.
Our aim is to provide and exemplify a working framework to test the BM or any other conjecture about the quantum nature of BH, provided quantitative predictions for the QNM spectrum, as in the FK model. 
We find that such tests are possible with current interferometric detectors at their design sensitivity, opening up the possibility to test quantum gravity effects with BH ringdown as theoretical predictions will become available in the upcoming future.

In Section~\ref{sec:BM conjecture_section} the BM conjecture and its application to the QNM spectrum -- according to the FK model -- is reviewed and discussed in detail. 
Section~\ref{sec:Methods section} outlines the analysis methods by describing the adopted waveform model and the Bayesian time-domain data analysis framework.
 Section~\ref{sec:GW150914 section} is devoted to the application of our method on GW150914~\cite{GW150914}, while in Section~\ref{sec:GWTC section} our analysis encompasses the population of GW events detected thus far by the LIGO-Virgo collaboration (LVC) for which a ringdown signal described by our model is observable.
Section~\ref{sec:Population section} presents a study performed on a simulated population of BBH signals. This study allows us to properly interpret the population results obtained on real events and to investigate the feasibility of conclusive tests on our working model with the current infrastructure of GW detectors. Finally, Section~\ref{sec:Summary_section} presents a summary of our results, together with future prospects.

\section{Bekenstein-Mukhanov conjecture: QNM spectrum}\label{sec:BM conjecture_section}

In this section we will introduce the basic idea of the BM conjecture, review the FK model, and clarify under which assumptions it can be applied to compute an alternative prediction of the QNM spectrum.
By BM conjecture we mean the hypothesis according to which the event horizon area of a non-extremal BH is quantised in units of the Planck length squared, proportionally to a dimensionless universal constant $\alpha \in \mathbb{R^+}$. More precisely~\cite{BEKENSTEIN1974, BEKENSTEIN-MUKHANOV},
\begin{equation}\label{eqn:area_conjecture}
A_H^{Q} = \alpha l_P^2 N = \alpha \frac{\hbar G}{c^2} N     \,,
\end{equation}
where $l_P \sim 1.6 \times 10^{-35}\,~\mathrm{m}^2$, $N \in \mathbb{Z}^+$, $\hbar = 1.05 \times 10^{-34}$ J s, $G = 6.67 \times 10^{-11} \mbox{N} \, \mbox{m}^2 \, \mbox{kg}^{-2}$, and $c = 2.99 \times 10^8 \, \mbox{m} \, \mbox{s}^{-1}$ (in the rest of this work we will use geometrised units in which $G = c=1$).
The BM proposal has received a lot of attention in the past thirty years ~\cite{BEKENSTEIN1973, MUKHANOV, HOD, MAGGIORE2008, VAGENAS, MEDVED, DAVIDSON, Coates_2019, ELMENOUFI2020} and several theoretical arguments have been advanced to calculate $\alpha$.

Even though the model is intrinsically heuristic (given the lack of a full quantum theory of gravity) and it is not clear how to interpret some of its features, e.g., the degeneracy of each area level~\cite{FOIT-KLEBAN}, or how it may deal with Hawking's information paradox and the unitarity problem~\cite{MATHUR-2009, ENTROPY-HAWKING2020}, this conjecture is still considered to be a useful starting point for the investigation of quantum features of BHs~\cite{Coates_2019, Chakraborty_2019, ELMENOUFI2020, AGULLO2020}, in that it basically involves well-understood principles and simple arguments from ordinary quantum mechanics. 
A definitive theoretical computation of $\alpha$ has not been currently attained and different values have been proposed in the literature, some of which being: $\alpha = 8 \pi$ \cite{BEKENSTEIN1973, MAGGIORE2008}, $\alpha = 4 \ln{2}$ \cite{MUKHANOV}, $\alpha = 4 \ln{3}$ \cite{HOD}, $\alpha = 8 \ln{2}$ \cite{DAVIDSON} (see~\cite{FOIT-KLEBAN} for a review on these different estimations). 
One of the compelling features of the BM model is that the value of $\alpha$ is universal, hence it is the same for any kind of BH regardless of its specific values of mass, angular momentum, or electric charge.

Following the intriguing analogy of ``quantum black holes as atoms'' \cite{BEKENSTEIN1996, BEKENSTEIN1997QUANTUM, BEKENSTEIN2015}, it is easy to derive the energy spectrum of such a QBH. The idea is to identify the quantum prediction for the outer horizon area with the classical one:
\begin{equation}\label{eqn:area_equality}
A_H^{Q} = A_H^{Kerr} \,,
\end{equation}
where $A_H^{Kerr}$ is the horizon area of a non-charged Kerr BH of mass $M$ and angular momentum $J$, 
\begin{equation}
A_H^{Kerr} = 8 \pi M^2 \Bigl( 1 + \sqrt{1 - a^2} \Bigr)  \,,
\end{equation} 
where $a \equiv J/M^2$ is the BH dimensionless spin ($0 \leq a \leq 1$).
From Eq.~(\ref{eqn:area_equality}), a minimal change in the area is given by $- \alpha \hbar (k+1)$, where~\footnote{For consistency with QNM literature, we shift the value of $k$ by one so that $k=0$ corresponds to the fundamental transition. One may redefine $k = -\Delta N$ without affecting the argument.} $(k+1) = - \Delta N$ is the change in the area quantum, with $k \geq 0$, whereas the total differential of the Kerr formula with respect to $M$ and $J$ gives 
\begin{equation}\label{eqn:Kerr_area_variation}
\Delta A_H^{Kerr} = \frac{16 \pi M}{\sqrt{1 - a^2}} \biggl[ \Bigl( 1 + \sqrt{1 - a^2} \Bigr) \Delta M - \frac{a}{2 M} \Delta J \biggr] \mbox{ .} 
\end{equation}
If we equate these quantities and invoke the principles of quantum mechanics for energy and angular momentum, $\Delta M = - \hbar \omega_k$ and $\Delta J = - \hbar m$ (where $m$ is the azimuthal quantum number of the quanta), we can derive:
\begin{equation}\label{eqn:quantised_frequencies}
\omega^{BM}_k = \frac{1}{M} \frac{(k+1) \alpha \sqrt{1 - a^2} + 8 \pi a m}{16 \pi (1 + \sqrt{1 - a^2})} \,, \quad k \geq 0  \mbox{ .}
\end{equation}
These are the frequencies that a BM QBH is allowed to emit and absorb~\cite{FOIT-KLEBAN,BEKENSTEIN2015}. Interestingly, Eq.~(\ref{eqn:quantised_frequencies}) implies that the QBH cannot absorb quanta below a certain threshold frequency. Though this may seem to preclude the possibility of having Hawking's (continuous) radiation, Bekenstein showed that by taking account of the quantum nature of the wave associated to the energy radiation there is no paradox at all~\cite{BEKENSTEIN1974}. 

Until now, there has not been an explicit measurement capable of proving or disproving this conjecture with data.
GWs may open up this possibility: under certain assumptions (see the section ``Testing Bekenstein-Mukhanov with LIGO'' of ~\cite{FOIT-KLEBAN}) the ringdown part of the  GW emitted by the coalescence of a BBH system could differ from its GR prediction, offering a way to test the conjecture. 
In the next sections we will show how this can be pursued by applying Bayesian inference to GW observations.
From here on, we will follow the theoretical model outlined by FK in~\cite{FOIT-KLEBAN}, where a heuristic interpretation of the quantised frequency formula, Eq.~(\ref{eqn:quantised_frequencies}), has been put forward in order to assess its validity using GW ringdown signals.

The main assumption of the FK model is that the area quantisation hypothesis modifies the boundary conditions at the horizon in a way that it also affects the light ring structure~\footnote{If the area quantisation hypothesis does not affect the light ring structure, there could still be some observable effects in the very last part of the ringdown through GW \emph{echoes}, as pointed out in~\cite{CARDOSO-FOIT-KLEBAN}.}. 
This is a necessary condition since the light ring structure determines the observed dominant ringdown frequency~\cite{GOEBEL1972, FERRARI-MASHHOON, CARDOSO-FRANZIN-PANI} that, as long as we consider the least-damped mode with small imaginary part, coincides with the QNM real part~\cite{MAGGIORE2008}. 
This assumption is guided by theoretical results showing that in several classes of hairy Schwarzschild BHs, hairs are allowed to show up above the photon sphere (black holes have no ``short'' hairs)~\cite{NUNEZ1996, HOD2011}; for axisymmetric BHs as the ones we are considering here, to the best of our knowledge, similar results have not been found, but attempts to resolve the information paradox have put forward compelling arguments in favour of the existence of modifications to the BH structure in the general case~\cite{Giddings:2017jts}.

Thus, in the FK model the value given by Eq.~(\ref{eqn:quantised_frequencies}) for a fundamental QBH transition ($k=0$ in our convention so that $|\Delta N|= 1$) matches the observed ringdown frequency: the QBH relaxes towards its ``fundamental'' state~\cite{FOIT-KLEBAN, BEKENSTEIN1974}. The huge number of transitions that may occur between different area levels only happen between adjacent states, where the amount of energy carried by each quantum is the same. 
The choice of a specific value of $k$ is not fine-tuned, as $k$ is degenerate with $\alpha$: the measured quantity is always the combination $k \alpha$, hence a choice of a specific value of $k$ would simply rescale the value of $\alpha$, the absolute value of which is currently unknown.
The value of $m$ depends on the helicity of the carrier of quanta, which for gravitons can be $m = \pm 2$. 
The physical picture of the process is that the QBH is emitting gravitons losing energy.
According to Hawking's area theorem~\cite{HAWKING1971-AREA, PENROSE-FLOYD} or to the Bekenstein-Hawking entropy law~\cite{BEKENSTEIN1973, HAWKING1974}, the surface area of the horizon of a BH can never decrease, generally increasing in a dynamical process.
As a consequence the right-hand-side of Eq.~(\ref{eqn:Kerr_area_variation}) \emph{has} to be positive:
$\Delta M < 0$ as the system is losing mass, so $\Delta J$ must be negative, leaving only $m=+2$ as possible choice.
The assumption underlying the Bekenstein-Hawking law is that the system is isolated, as it is reasonable to assume for the merging binary systems that we will consider. 
Based on a classical description, the $m = -2$ polarization could be still possible if we had absorption of angular momentum, as it happens in case of superradiance~\cite{ZELDOVICH1971, MISNER1972}, i.e. stimulated emission, that however in an isolated system cannot be present.

The outcome of these considerations is the following relation for the quantum QNM fundamental frequency:
\begin{equation}\label{eqn:quantum_QNM_fundamental_freq}
\omega_0 (M_f, a_f, \alpha) = \frac{1}{M_f} \frac{\alpha \sqrt{1 - a_f^2} + 16 \pi a_f}{16 \pi (1 + \sqrt{1 - a_f^2})}   \mbox{ ,}
\end{equation} 
where we now refer to the final mass $M_f$ and final spin $a_f$ of the remnant BH resulting from a BBH merger.
We note that the frequency $\omega_0$ is a continuous function of its variables, meaning that $M_f$, $a_f$, and $\alpha$ can vary freely between their allowed positive domains. 

A way to interpret Eq.~(\ref{eqn:quantum_QNM_fundamental_freq}) is that of a law that takes as input the parameters of the QBH and outputs its fundamental characteristic oscillation frequency, in a similar fashion to GR, where numerical relativity fits for the real part of QNMs give the frequency of the ringdown signal for a given $M_f$ and $a_f$~\cite{LISA_spectroscopy}.
Notably, given a pair of $M_f$ and $a_f$, there is no limiting value of $\alpha$ for which this relation can  predict the whole spectrum of frequencies of the GR fundamental QNM~\cite{FOIT-KLEBAN, LISA_spectroscopy}; this implies that if a GR signal is analysed using a QBH ringdown template based on $\omega_0$, in general one should expect to recover a value of $\alpha$ that will depend on those specific event parameters, that is, the value will change event by event due to its correlation with the other BH parameters.

Caveats regarding Eq.~(\ref{eqn:quantum_QNM_fundamental_freq}) mainly stem from the fact that one cannot predict the time of emission of gravitons (although a lower bound on the decay rate can be estimated~\cite{AGULLO2020}).
Nonetheless, given the assumptions on the light ring previously discussed, in the FK approach Eq.~(\ref{eqn:quantum_QNM_fundamental_freq}) is interpreted as a predictive law for the ringdown oscillation frequency. 
Even though it is quite improbable that single gravitons could be measured, to support this view it may be helpful to think of a GW as a coherent superposition of a large number of gravitons~\cite{DYSON, PARICK-WILCZEK-ZAHARIADE2020}. 
Stated alternatively, gravitons in a GW are phase-coherent~\cite{Flanagan2005}, meaning that the frequency is the same for all of them: the remnant QBH should be a coherent source as well, as the gravitational radiation is emitted by a single coherent gravitational structure. 
By analogy of the QBH as a ``hydrogen atom of quantum gravity''~\cite{BEKENSTEIN2015}, problems of incoherence may arise when numerous atoms radiate together incoherently. This case is excluded as a system like a remnant BH is made of a single source (atom). Also, the radiation is a stimulated emission, since in the context of a BBH coalescence the final BH emits radiation under the influence of the excitation produced during the merger phase. The fundamental frequency, Eq.~(\ref{eqn:quantum_QNM_fundamental_freq}), remains numerically the same for a considerable range of values of $M$ for fixed $a$. So during the process of GW emission, although the remnant BH will lose mass, it would still be emitting at the same frequency range of a certain width.
In this scenario, the ringdown signal will differ from the classical prediction, thus offering the possibility to test the FK proposal.


\section{Methods}\label{sec:Methods section}

\subsection{Waveform models}\label{subsec:Waveform models}

We assume a ringdown waveform model of the form:
\begin{equation}\label{eqn:Kerr_template}
h_+ - i h_{\times} =
 \frac{M_f}{D_L} \left( \tilde{\mathcal{A}}\, e^{i (t - t_{0})\tilde{\omega}} + \tilde{\mathcal{A}^{'}}  e^{-i (t - t_{0})\tilde{\omega}^*} \right) \,,
\end{equation}
where the complex amplitudes are generally written as $\tilde{\mathcal{A}} = \mathcal{A} e^{i \phi}$, the QNM frequencies have the form $\tilde{\omega} = \omega + i/\tau$, where $\omega$ and $\tau$ in general may depend on $M_f$ and $a_f$, the detector-frame mass and dimensionless spin of the remnant BH, $D_L$ is the luminosity distance (measured in Mpc) to the source, and $t_{0}$ is the start time of the ringdown, that following the work in~\cite{O3a-TGR-PAPER} will be assumed to be $t_0 = t_\mathrm{{peak}} + 10 M_f$, where $t_\mathrm{{peak}}$ is the peak of the complex strain, ensuring that a linearised description using a single mode is valid~\footnote{The FK model gives an interpretation of the dominant quadrupolar GW emission only. Hence, in this study multi-mode ringdown signals will not be considered.
}.
The two terms in the brackets correspond to the two conjugate pairs of solutions appearing in the Teukolsky equation~\cite{LISA_spectroscopy}.

On one hand, in case of a ringdown signal emitted by a classical GR BH the complex frequency $\tilde{\omega}$ corresponds to the longest-lived mode: we label this waveform model \emph{GR}. 
On the other hand, the FK model assigns the real part of the \emph{quantum} fundamental QNM, Eq.~(\ref{eqn:Kerr_template}), while it does not predict its imaginary part. 

In order to have a testable waveform model, we need to define a \emph{quantum} damping time to associate to $\omega_0$, which we henceforth label $\tau_0$.
It is important to note that, to the best of the authors' knowledge, no precise prediction is available in the literature for the quantum fundamental damping time.
As a consequence, we may treat $\tau_0$ as a free independent parameter to be inferred from the data: we will refer to this model as $nGR_F$, the $F$ standing for a free $\tau_0$. Alternatively, we may assume a functional relation involving the QBH parameters, meaning that $\tau_0$ will depend on the BH mass, spin, and $\alpha$ through an ansatz. 
The dissipative nature of the ringdown process suggests the introduction of the \emph{quality factor} $Q \equiv \pi f \tau$, which relates the frequency $f$ and damping time $\tau$ of a given QNM \cite{LISA_spectroscopy}. 
Arguing that the damping time must depend on the parameter $\alpha$, we may assign $\tau_0 (M_f, a_f, \alpha) = \tau_0(\omega_0, Q_{\textup{GR}}) $, using the quantised frequency in Eq.~(\ref{eqn:quantum_QNM_fundamental_freq}) along with $Q_{\textup{GR}}$, the quality factor of the longest-lived QNM in GR:
\begin{equation}\label{eqn:quantum_QNM_fundamental_tau}
\tau_0 (M_f, a_f, \alpha) = 2 \frac{Q_{\textup{GR}}(a_f)}{\omega_0(M_f, a_f, \alpha)} \mbox{ .}
\end{equation}
We will dub the model assuming this relation $nGR_A$, where the $A$ reminds us that we are assuming an ansatz for $\tau_0$.
It may be argued that we are using a \emph{classical} quality factor in the definition of a \emph{quantum} damping time: in Sec.~\ref{sec:GW150914 section} we will use real data to explore the reasonableness of our guess and compare the values predicted by Eq.~(\ref{eqn:quantum_QNM_fundamental_tau}) with the distribution of $\tau_0$ obtained from an analysis where we directly sample on it.
This will help us in defining the working hypothesis for our alternative ringdown model that we will use in the rest of our study.

To summarize, in the next sections we will be working with the following ringdown waveform models: 
\begin{itemize}
\item[i)] $GR$, defined by Eq.~(\ref{eqn:Kerr_template}) with $\tilde{\omega}$ as predicted by GR; 
\item[ii)] $nGR_F$, defined by Eq.~(\ref{eqn:Kerr_template}) with $\omega = \omega_0 (M_f, a_f, \alpha)$ and $\tau$ left as a free parameter $\tau_0$;
\item[iii)] $nGR_A$, defined by Eq.~(\ref{eqn:Kerr_template}) with $\omega = \omega_0 (M_f, a_f, \alpha)$ and $\tau = \tau_0 (M_f, a_f, \alpha)$. 
\end{itemize}

\begin{figure}[h]
\center
\includegraphics[width=0.9\textwidth]{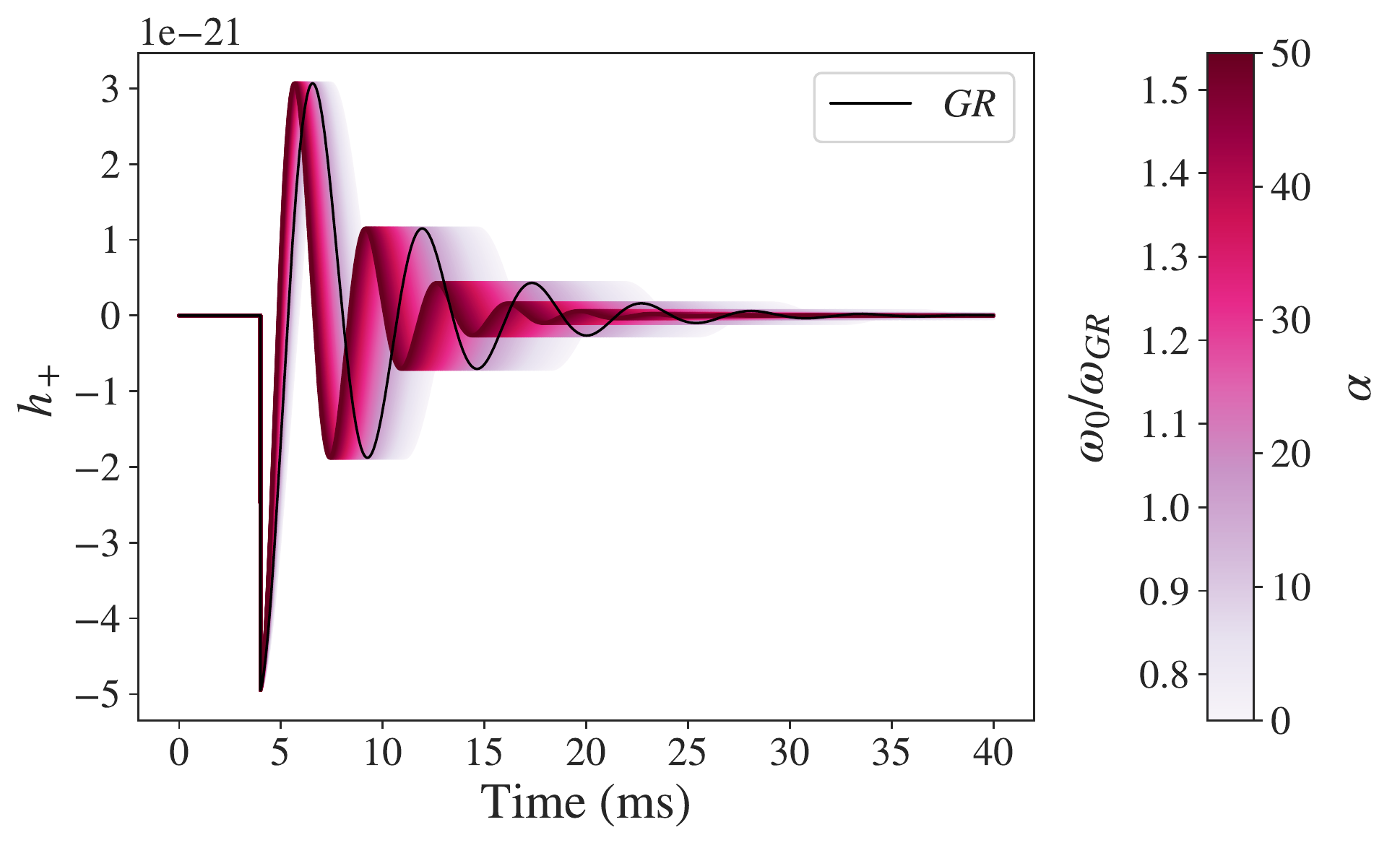}
\caption{$GR$ ringdown waveform (black line) and $nGR_A$ ringdown waveforms (scaled from white to red) for different values of $\alpha$ and $\omega_0/\omega_{GR}$, where $\omega_{GR}$ is the frequency associated with the GR waveform. See the text for the waveform definitions.}
\label{fig:3D_wf_alpha} 
\end{figure}

To help visualize how a QBH ringdown waveform may look compared to a $GR$ one, 
in Fig.~\ref{fig:3D_wf_alpha} we plot the ``+'' polarization of the $GR$ and $nGR_A$ waveforms as a function of time, given a common set of parameters characterizing the source ($D_L = 413.1$ Mpc, $\mathcal{A} = \mathcal{A}' = -0.91$, $\phi = 0$, $M_f = 91.6 M_{\odot}$, $a_f = 0.68$): the $GR$ waveform is represented by the black curve, while the $nGR_A$ waveform is shown for increasing values of $\alpha$ from white ($\alpha=0$) to red ($\alpha=50$).
We can now visually appreciate the differences between the two template families,  i.e., the lack of a value of $\alpha$ for which a QBH may ``mimic'' a classical BH. Note that given $M_f$ and $a_f$, there can be a value of $\alpha$ such that the frequency is equal to that of GR: the waveforms in the panel show that there is no couple of frequency \emph{and} damping time that exactly mimics those of GR. Moreover, it is relevant to note that higher values of $\alpha$ tend to produce higher ringdown frequencies (from Eq.~(\ref{eqn:quantum_QNM_fundamental_freq})), hence smaller damping times (from Eq.~(\ref{eqn:quantum_QNM_fundamental_tau})).

\subsection{Bayesian data analysis}\label{subsec:Bayesian data analysis}

Given a ringdown signal and a waveform template, we can: (i) estimate the parameters of the BH emitting the signal; (ii) compare different theories modeling the emission. Both these tasks are naturally accomplished within the framework of Bayesian data analysis~\cite{GREGORY}.

Bayesian probability theory describes the state of knowledge about an uncertain hypothesis $\mathcal{H}_i$, given our prior information $I$, as a probability, labeled $p(\mathcal{H}_i\,|\,I) \in [0,1]$ (to ease the notation sometimes the $I$ will be omitted). 
Probabilities on continuous variables (e.g. the parameters describing a BBH coalescence $\vec{\theta} = \{ \theta_1, \ldots, \theta_N \}$) are described via a probability density function (PDF), $p(\vec{\theta} \,|\, \mathcal{H}_i, I)$, where $\int \! \textup{d} \theta \, p(\vec{\theta} \,|\, \mathcal{H}_i, I) = 1$.

Estimates of parameters then follow directly from Bayes theorem~\cite{REVEREND}, where a \emph{prior} PDF $p(\theta\,|\,\mathcal{H}_i, I)$ is updated upon receiving the new data $D$ from the experiment to give a \emph{posterior} PDF $p(\vec{\theta} \,|\, D, \mathcal{H}_i, I)$:
\begin{equation}
p(\vec{\theta} \,|\, D, \mathcal{H}_i, I) = \frac{p(\vec{\theta} \,|\, \mathcal{H}_i, I) \,p(D \,|\, \vec{\theta}, \mathcal{H}_i, I)}{p(D \,|\, \mathcal{H}_i, I)} \,,
\end{equation}
where $p(D \,|\, \vec{\theta}, \mathcal{H}_i, I)$ is the \emph{likelihood function} for the observation $D$, assuming a given value of the parameters $\vec{\theta}$ and the model $\mathcal{H}_i$. The denominator, called \emph{evidence}, is simply the expectation value of the likelihood 
function over the prior:
\begin{equation}
p(D \,|\, \mathcal{H}_i, I)  = \! \int \!\! \textup{d} \theta_1 \! \cdots \textup{d} \theta_N \, p(D \,|\, \vec{\theta}, \mathcal{H}_i, I) \,p(\vec{\theta} \,|\, \mathcal{H}_i, I) \mbox{ .}
\end{equation}
Results for specific parameters are found by marginalising the multi-dimensional posterior distribution over the unwanted parameters:
\begin{equation}
p(\theta_1 \,|\, D, \mathcal{H}_i, I) = \int \! \textup{d}\theta_2 \cdots \textup{d}\theta_N \, p(\vec{\theta} \,|\, D, \mathcal{H}_i, I) \mbox{ .}
\end{equation}
We can then use these marginal PDFs to extract any estimate that is needed, e.g. median values or credible regions (CRs).

Now, say we want to perform model selection between two competing hypotheses, i.e. compare two competing (mutually exclusive) models $\mathcal{H}_i$ and $\mathcal{H}_j$, in light of some made observations. 
For instance, they could assume two different waveform templates and we would like to see which one better matches the data. 
We can compute the ratio of posterior probabilities, known as the \emph{odds' ratio},
\label{eqn:odds_ratio}
\begin{eqnarray}
O^i_j &= \frac{p(\mathcal{H}_i\,|\, I)}{p(\mathcal{H}_j\,|\, I)} \frac{p(D \,|\,\mathcal{H}_i , I)}{p(D \,|\, \mathcal{H}_j, I)} \nonumber \\  
&=  \frac{p(\mathcal{H}_i\,|\, I)}{p(\mathcal{H}_j\,|\, I)} \,\mbox{B}^i_j \,,
\end{eqnarray}
where B$^i_j$ is called \emph{Bayes factor}, the central quantity used in model selection studies.

Consider for simplicity a single detector: the data $d$ consist of a time series sampled at discrete times. The data can be modelled as the sum 
of the strain $h$ of the ringdown waveform, Eq.~(\ref{eqn:Kerr_template}), and a noise component $n$,
\begin{equation}
d(t) = h(t) + n(t) \mbox{ .} 
\end{equation} 
Our waveform model $h$ can be thought of as a function that takes as input some parameters and produces as output $h_{+,\times}$ in the time-domain. More precisely, what we measure is a linear combination of the polarizations, called \emph{strain}, defined as $h = F_+(\alpha', \delta', \psi) h_+ + F_{\times}(\alpha', \delta', \psi) h_{\times}$, where the functions $F_{+,\times}$ are the \emph{antenna pattern functions} of the detectors~\cite{HAWKING300YEARS, MAGGIORE-GW1}.

The parameter vector $\vec{\theta}$ includes the so-called intrinsic and extrinsic parameters: final mass $M_f$, final spin $a_f$, mode amplitudes $\mathcal{A}$, $\mathcal{A}'$ and phases $\phi$, $\phi'$, the luminosity distance to the source $D_L$, sky location parameters like right ascension $\alpha'$ and declination $\delta'$ of the source~\footnote{We use primed labels for the angles to avoid possible confusion between the right ascension $\alpha'$ and the QBH universal constant $\alpha$.}, polarization angle $\psi$ describing the orientation of the projection of the binary's orbital momentum vector onto the plane on the sky~\cite{ANDERSON-BRADY-CREIGHTON-FLANAGAN}, the inclination $\iota$ between the line of sight and the angular momentum vector of the source and the start time of the ringdown $t_0$.

\subsection{Ringdown time-domain analysis}\label{subsec:Ringdown time-domain analysis}

Our analysis is performed in the time-domain using data from the Advanced LIGO detectors, provided by the Gravitational-Wave Open Science Center~\cite{GWOSC}.
The detector noise is modelled as a wide-sense stationary Gaussian process. The validation of this assumption requires a careful analysis of the noise (see e.g.~\cite{LVC-NOISE} for an application of the validation process to GW150914 and~\cite{LVC-GUIDE} for a general review on detector-noise analysis and signal extraction). The stochastic process describing the detector noise is then completely characterized by its two-point auto-covariance function $C(\tau)$:
\begin{equation}
C(\tau) = \int{\! \textup{d}t \, n(t) \, n(t+\tau)} \,,
\end{equation}
which we estimate from a segment of data surrounding the event. 
When analyzing LIGO and Virgo interferometric data, sampled at a rate of 4096~Hz, we apply a band-pass 4th order Butterworth filter in the band [20,2043] Hz.
The data are subsequently split into $4$-seconds long chunks. 
The auto-covariance function is computed as the ensemble average of the individual auto-covariances estimated on each chunk~\cite{BENDAT-PIERSOL}, excluding the one containing the time of the peak of $h_+^2 + h_{\times}^2$, that is used as reference time to measure the onset of the ringdown regime.
For the simulated data considered in this work, the auto-covariance function is estimated by using the Wiener-Khinchin theorem~\cite{BENDAT-PIERSOL}, hence inverse Fourier transforming the predicted power spectral density of the LIGO and Virgo detectors at their design sensitivity~\cite{LIGO_VIRGO_DESIGN_SENS}.
The time-domain log-likelihood function for the observed strain series $d(t)$, given the presence of 
a GW signal $h(t; \vec{\theta})$ is:
\begin{equation}\label{eqn:likelihood_function}
\log p(d|\vec{\theta} ,I) \! = \! -\frac{1}{2} \! \int \! \! \int \!{\textup{d}t\,\textup{d}\tau \, r(t;\vec{\theta}) \, C^{-1}(\tau) \, r(t\!+\!\tau ; \vec{\theta})} \,, 
\end{equation}
where $r(t; \vec{\theta}) = d(t) - h(t;\vec{\theta})$ is the residual and the domains of integration extend over $[t_0, t_0+0.1\,\mbox{s}]$. In case of simulated data, we inject the signals into zero-noise, so that our results are independent of a specific noise realization.
When considering a network of interferometric detectors, since in the absence of a GW signal the individual data streams are 
independent, the joint likelihood is given simply by the product of the likelihoods of the single detectors. We perform the analysis using \emph{pyRing}~\cite{CARULLO-DELPOZZO-VEITCH, Isi:2019aib, O3a-TGR-PAPER}, a time-domain parameter estimation package based on a nested stochastic sampling algorithm~\cite{DELPOZZO-VEITCH, SKILLING2006} and tailored to specifically analyse ringdown signals.

\subsection{Prior choices}\label{subsec:Prior_choices}

Unless otherwise specified, for the analysis of interferometer data we assume uniform priors as follows:
$\psi \in [0,\pi]$, $\cos \iota \in [-1,+1]$, $a_{f} \in [0,0.99]$, $M_{f} \in [10,500] M_{\odot}$, $\mathcal{A},\mathcal{A}' \in [0,50]$, $\phi,\phi' \in [0, 2\pi]$, $\alpha \in [0, 50]$, while we assume a (quadratic) uniform comoving distance distribution with $\log D_L \in [\log 10, \log 10^4]$; the sky position is fixed to the maximum likelihood value inferred from an inspiral-merger-ringdown (IMR) analysis of the event~\cite{O3a-TGR-PAPER, O3a-CBC-CATALOG}, and $t_0$ is fixed to $10 \; M_f$ after the peak of $h_+^2 + h_{\times}^2$, computed using median values again fixed by the corresponding IMR analysis. When sampling on $\tau_0$, we assume a uniform prior $\tau_0 \in [0.5, 50]$ ms.  In case of simulated data, we will instead sample over the entire parameter space assuming prior distributions in agreement with those used to generate the population parameters, that will be discussed in detail in Sec.~\ref{sec:Population section}. 
Finally, in the remainder of the paper we quote parameter estimates as median and 90$\%$ CRs.

\section{Inferring $\alpha$ from GW150914}\label{sec:GW150914 section}

GW150914 is the first BBH coalescence event detected by the two LIGO instruments on September 14, 2015~\cite{GW150914}. 
The signal was analysed using accurate signal models developed under the assumption that GR is the underlying theory of gravity. 
The analysis concluded that GW150914 was generated by the coalescence of two BHs of detector-frame masses $39^{+5}_{-4} M_{\odot}$ and $33^{+4}_{-5} M_{\odot}$ at a luminosity distance of $440^{+160}_{-180}$ Mpc which formed a remnant BH of detector-frame mass $68^{+4}_{-4} M_{\odot}$ and final spin $0.68^{+0.05}_{-0.06}$~\cite{LVC-GW150914-PRX}.

The ringdown of GW150914 has already been thoroughly studied to extract remnant properties and produce accurate tests of GR~\cite{TGR-LVC2016, CABERO2018, Brito:2018rfr, CARULLO-DELPOZZO-VEITCH, Isi:2019aib, CalderonBustillo:2020tjf}.
Our aim is to use this event, which is the loudest so far, as a test-bed to quantify how the inference changes assuming either $nGR_F$, $nGR_A$, or $GR$ as our inference model. At the same time, motivated by the analysis reported in \cite{FOIT-KLEBAN}, we want to assess whether GW150914 can provide constrains on the parameter $\alpha$ that enters the area conjecture formula, Eq.~(\ref{eqn:area_conjecture}). 

We test the models defined in Sec.~\ref{subsec:Waveform models} applying the formalism presented in Sec.~\ref{subsec:Bayesian data analysis} to infer all the parameters $\vec{\theta}$ of the waveform. 
We sample over the parameters adopting the techniques detailed in Sec.~\ref{subsec:Ringdown time-domain analysis} and assuming the prior distributions reported in Sec.~\ref{subsec:Prior_choices}; in particular, we remark that our choice of prior distribution on $\alpha$ encompasses all the values that are considered to be reasonable from theoretical estimates~\cite{CARDOSO-FOIT-KLEBAN}.

\subsection{$nGR_F$ model}

A first ``agnostic'' measure of $\alpha$ can be obtained by making the minimal number of assumptions: this corresponds to adopt the $nGR_F$ waveform model introduced in Sec.~\ref{subsec:Waveform models}, which prescribes to measure the theoretically-unknown $\tau_0$ along with all the other parameters.   
This is the most conservative assumption we can make in absence of quantum-motivated predictions on the damping time, although we are in principle losing some predictive power by including an additional parameter.

We will now show results on the inference of $\alpha$ according to two possible sets of priors on the remnant mass and spin. 
In both cases all the measured extrinsic and intrinsic parameters agree with the full IMR analysis of the signal made by the LIGO-Virgo Collaboration (LVC)~\cite{LVC-GW150914-PRX}.

\subsubsection{Uninformative priors for $M_f$ and $a_f$}  \quad

\begin{figure}[h]
\center
\includegraphics[width=0.7\textwidth]{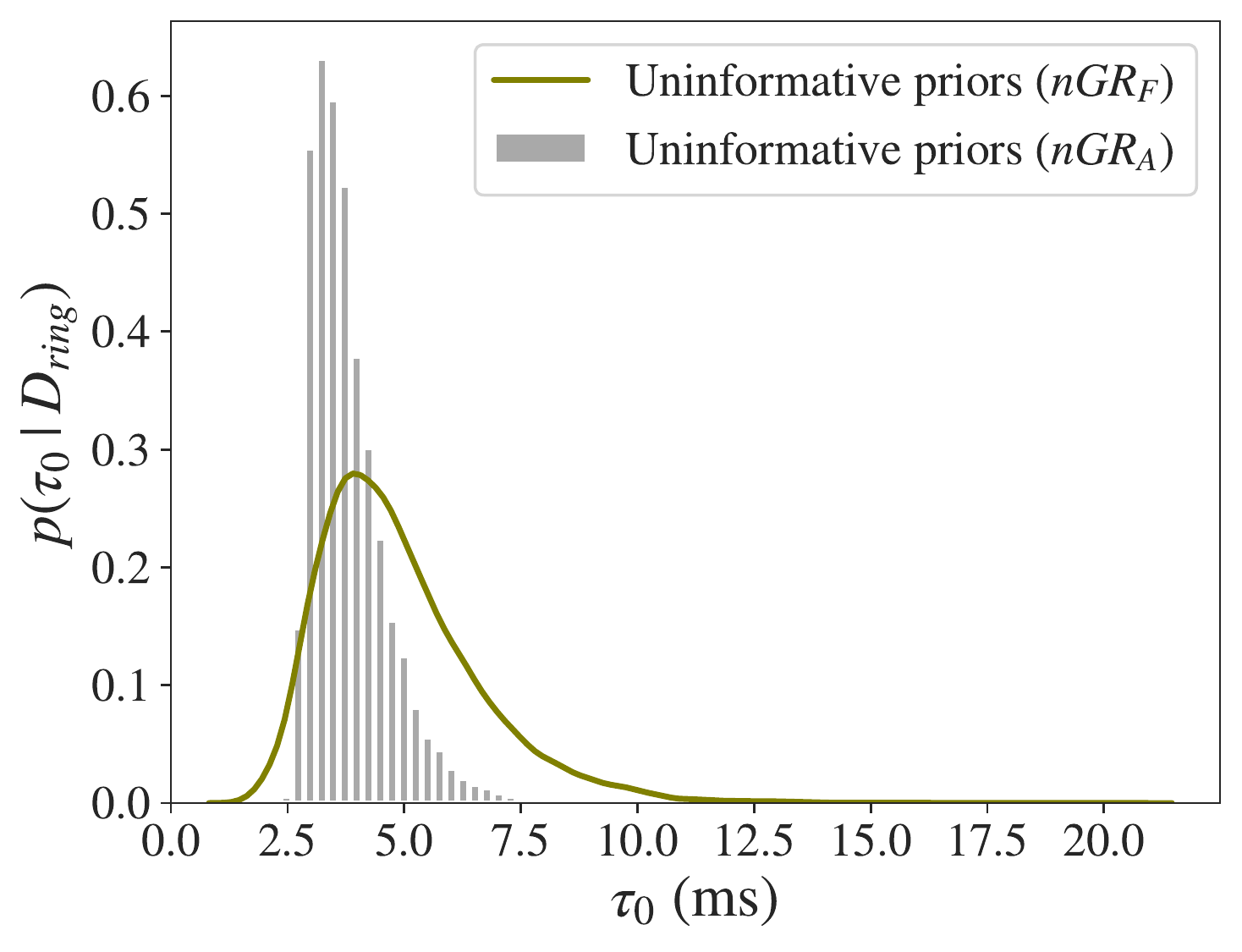}
\caption{One-dimensional posterior distribution of $\tau_0$ (olive line) inferred from GW150914 ringdown data $D_{ring}$ assuming uninformative priors with $\tau_0 \in [0.5, 50]$ ms: recovered value $\tau_0 = 4.6^{+3.6}_{-1.8}$ ms. By comparison, the dark grey histogram (recovered value: $3.6^{+1.8}_{-0.7}$) shows the values of $\tau_0$ as predicted by Eq.~(\ref{eqn:quantum_QNM_fundamental_tau}) using the values of the sampled parameters $M_f$, $a_f$, and $\alpha$.}
\label{fig:tau0_GW150914}
\end{figure}

\begin{figure}[h]
\center
\includegraphics[width=0.7\textwidth]{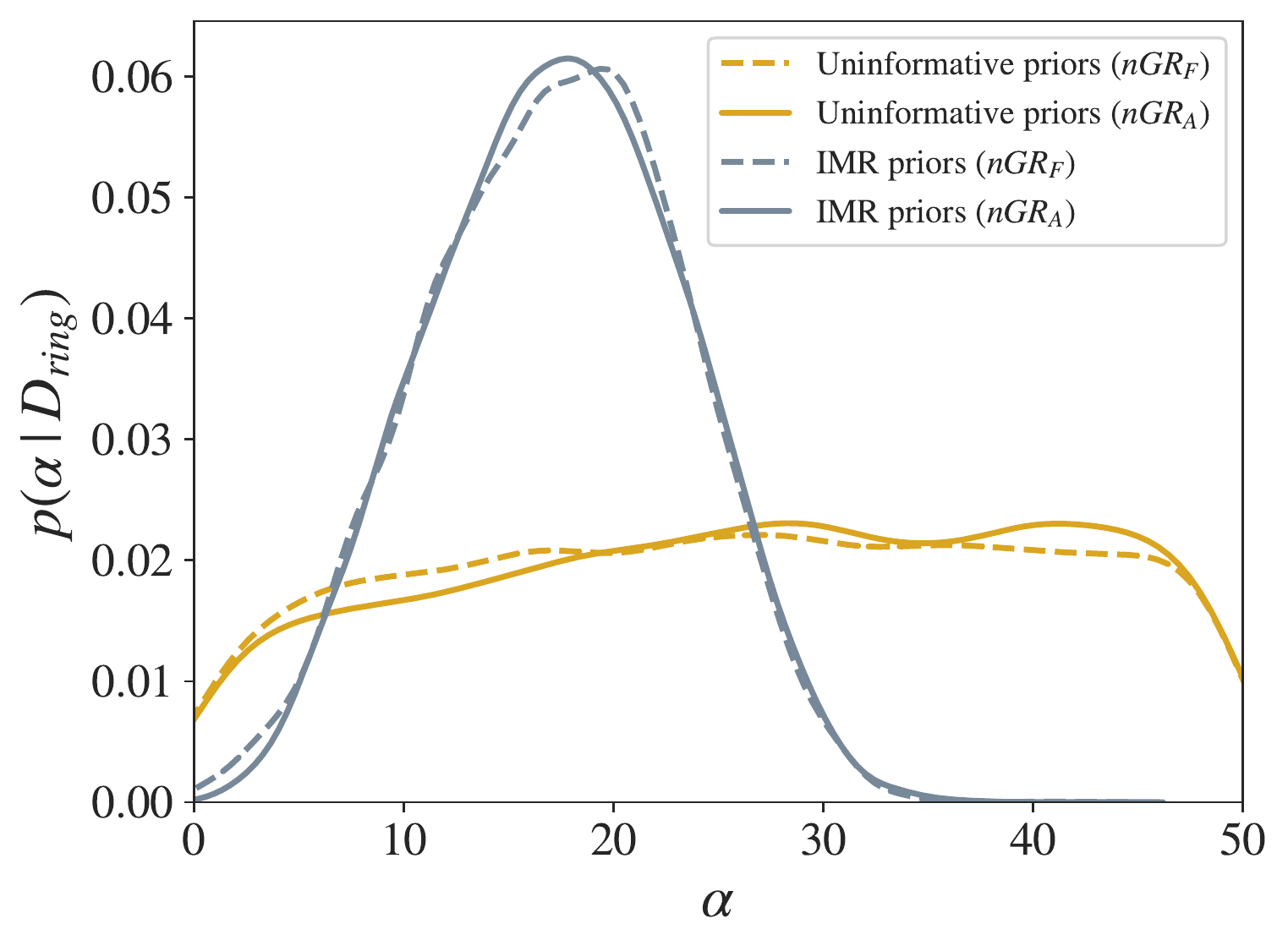}
\caption{Posterior distribution for the $\alpha$ parameter inferred from GW150914 ringdown data $D_{ring}$ assuming: a) $nGR_F$ with uninformative uniform priors: $M_f \in [10, 250] M_{\odot}$, $a_f \in [0.0, 0.99]$ (gold dashed line); b) $nGR_F$ with IMR uniform priors: $M_f \in [62, 75] M_{\odot}$, $a_f \in [0.55, 0.75]$ (grey dashed line); c) $nGR_A$ with uninformative uniform priors: $M_f \in [10, 250] M_{\odot}$, $a_f \in [0.0, 0.99]$ (gold solid line); d) $nGR_A$ with IMR uniform priors: $M_f \in [62, 75] M_{\odot}$, $a_f \in [0.55, 0.75]$ (grey solid line).}\label{fig:alpha_plot_GW150914}
\end{figure}

The most conservative choice of priors, which we call \emph{uninformative}, prescribes uniform intervals $M_f \in [10, 500] M_{\odot}$ and $a_f \in [0, 0.99]$. 
The posteriors of final mass and spin are reported in Table~\ref{tab:logBF}, where they can be compared with those obtained assuming $GR$ as the correct theory describing the ringdown emission. The former show a broadening with respect to the latter: this is consistent with 
the smaller amount of SNR present in the ringdown-only signal as well as with
the addition of extra parameters, which in turn decreases the information content that can be used by the algorithm to constrain the mass and spin.
The measure of $\tau_0$ is reported in Fig.~\ref{fig:tau0_GW150914}, where the olive line reports the marginalized one-dimensional PDF with $\tau_0 = 4.6^{+3.6}_{-1.8}$ ms. We will make additional comments on this measure when discussing the $nGR_A$ analysis, where $\tau_0$ will be predicted by the ansatz Eq.~(\ref{eqn:quantum_QNM_fundamental_tau}).
The posterior distribution for $\alpha$ does not differ appreciably with respect to the prior ($\alpha = 26.3^{+21.2}_{-22.8}$) as shown in Fig~\ref{fig:alpha_plot_GW150914} (gold dashed line), meaning that the signal is too weak to provide an informative measurement. In fact, as we will see in Sec.~\ref{sec:Population section}, a better strategy is to combine the information from multiple events, because single events -- even if moderately loud -- in general do not allow for a measure of $\alpha$ and, more importantly, for a validation of the measure. 
This conclusion differs from~\cite{FOIT-KLEBAN}, where a quite stringent bound $14 < \alpha < 18$ is reported. 
The difference in results can be understood by noting that the task of extracting all the parameters -- including $\alpha$ -- from interferometric detector data does not rely exclusively on accurate theoretical predictions of the GW signal, but also crucially on the employed analysis formalism (for an introduction to GW data analysis see~\cite{Flanagan2005}). 
The study in~\cite{FOIT-KLEBAN} constrains $\alpha$ \emph{a posteriori}, without a direct sampling of the parameter probability distribution; as a consequence, the uncertainties on all the parameters and the correlations between $\alpha$, $M_f$, and $a_f$ have not been fully taken into account. 
This explains the very stringent bound therein reported, which does not hold when accounting for the full parameter space. 
Here we instead employ a proper Bayesian data analysis framework which makes a minimum number of assumptions and infers $\alpha$ in concert with all the other relevant parameters. The posterior for $\alpha$ is then obtained by marginalizing the multidimensional posterior PDF over all the other parameters.

\begin{table}
\centering
\caption{Summary of the signal-to-noise Bayes factors, $M_f, a_f$ median and 90$\%$ CRs obtained with different priors for the final mass and spin. Each row corresponds to a different analysis of GW150914, assuming a signal as predicted by the waveform model reported in the first column. The statistical errors on the log Bayes factors are $\pm \, 0.1$.}
\begin{indented}
\item[] \begin{tabular}{@{}cccc}
\hline\hline
\multicolumn{4}{c}{GW150914} \\
\hline\hline
\multicolumn{4}{c}{Uninformative priors} \\
\hline
Model & logB$^s_n$ & $M_f/M_\odot$ & $a_f$\\
\hline
$nGR_F$   & $44.3$ &   $	67.7^{+\, 43.2}_{-\, 38.4}$	   &  $0.51^{+\, 0.40}_{-\, 0.45}$	 \\
$nGR_A$     & $46.7$ &   $70.4^{+\, 33.6}_{-\, 35.9}$	   &  $0.53^{+\, 0.31}_{-\, 0.42}$	 \\
$GR$ 		  & $46.8$ & $62.7^{+\, 19.0}_{-\, 12.1}$      &    $0.52^{+\, 0.33}_{-\, 0.44}$	   \\
\hline\hline 
\multicolumn{4}{c}{IMR priors} \\
\hline
Model & logB$^s_n$ & $M_f/M_\odot$ & $a_f$\\
\hline
$nGR_F$   & $46.9$ & $68.4^{+\, 6.0}_{-\, 5.8}$   & $0.65^{+\, 0.09}_{-\, 0.09}$ \\
$nGR_A$   & $49.5$ & $68.4^{+\, 5.9}_{-\, 5.8}$   & $0.65^{+\, 0.09}_{-\, 0.09}$ \\
$GR$   & $50.9$ & $68.1^{+\, 5.3}_{-\, 4.8}$ & $0.65^{+\, 0.09}_{-\, 0.09}$ \\
\hline\hline
\end{tabular}
\end{indented}
\label{tab:logBF}
\end{table}

\subsubsection{IMR priors for $M_f$ and $a_f$} \quad

We also explored a more constraining choice of priors, which we call \emph{IMR}, where the uniform distributions of $M_f$ and $a_f$ now broadly contain the 99$\%$ CRs of their one-dimensional PDFs obtained from a full IMR analysis of GW150914~\cite{LVC-GW150914-PRX}: $M_f \in [62, 75] M_{\odot}$ and $a_f \in [0.55, 0.75]$.
This choice corresponds to the case where $\alpha$ is weakly correlated with other BH parameters, assuming that prior to the ringdown phase the GW emission is dominated by classical effects as predicted by GR.
In this scenario, the emission of energy and angular momentum is correctly described by classical GR and as such the remnant mass and angular momentum will be approximately the ones estimated from a standard analysis. 
While we get a similar measure of the damping time, $\tau_0 = 4.5^{+3.2}_{-1.8}$ ms, in this case $\alpha$ is constrained much better compared to the uninformative set of priors. 
This is due to a a large fraction of the BH parameter space being excluded \emph{a priori} as a consequence of the stringent bounds imposed on the mass and spin of the remnant. 
The $\alpha$ posterior distribution is shown in Fig.~\ref{fig:alpha_plot_GW150914} (grey dashed line), with recovered value $\alpha = 17.4^{+9.5}_{-10.4}$.
This result is more easily comparable with that of~\cite{FOIT-KLEBAN}, while still adopting a robust framework for the analysis of the data, as previously explained. 
Differently from~\cite{FOIT-KLEBAN}, even with more stringent priors the posterior of $\alpha$ encloses all the theoretical proposals reported in Sec.~\ref{sec:BM conjecture_section}.
It is important to note that although this prior choice might seem reasonable, it is driven by very strong assumptions about the smallness of the putative modifications due to the non-GR dynamics prior to the linearised ringdown regime, neglecting most of the correlation structure between $\alpha$ and the Kerr BH parameters $M_f$ and $a_f$. 

\subsection{$nGR_A$ model}\label{subsec:nGR_A model}

We now focus on the $nGR_A$ model introduced in Sec.~\ref{subsec:Waveform models}, where the quantum damping time $\tau_0$ is assigned according to the ansatz given by Eq.(\ref{eqn:quantum_QNM_fundamental_tau}). We will compare the results with those obtained assuming the $nGR_F$ model.  
In this case as well, all the measured extrinsic and intrinsic parameters agree with the full IMR analysis of the signal reported by the LVC~\cite{LVC-GW150914-PRX}. We proceed as before with two alternative analyses based on uninformative and IMR priors for $M_f$ and $a_f$.

\subsubsection{Uninformative and IMR priors for $M_f$ and $a_f$}\quad

Estimates of the remnant parameters for the $nGR_A$ model in case of uninformative and IMR priors are reported in Table~\ref{tab:logBF}, showing CRs similar to the $nGR_F$ model.
First, in case of uninformative priors, it is interesting to compare the PDF of $\tau_0$, obtained in the $nGR_F$ analysis, with the values of $\tau_0$ predicted by Eq.~(\ref{eqn:quantum_QNM_fundamental_tau}) using the sampled values of $M_f$, $a_f$, and $\alpha$, obtained in the $nGR_A$ analysis. 
These latter are shown in Fig.~\ref{fig:tau0_GW150914} (dark grey histogram) along with the $nGR_F$ posterior of $\tau_0$.  It is important to note that the histogram does not represent a sampled quantity, but the values predicted by Eq.~(\ref{eqn:quantum_QNM_fundamental_tau}) (which makes use of sampled parameters). 
The inferred value of $\tau_0$ from the histogram ($3.6^{+1.8}_{-0.7}$) is consistent with the 90\% CRs of the PDF, suggesting that Eq.~(\ref{eqn:quantum_QNM_fundamental_tau}), hence the $nGR_A$ model, is a fairly reasonable working hypothesis.
The $\alpha$-posterior for both the uninformative and IMR priors are shown in Fig.~\ref{fig:alpha_plot_GW150914} (gold and grey solid lines, respectively). 
The $nGR_A$ and $nGR_F$ posteriors are very similar: with uninformative priors we get $\alpha = 27.4^{+20.1}_{-23.8}$, while with IMR priors we constrain $\alpha = 17.4^{+9.6}_{-10.1}$, with no appreciable difference from the $nGR_F$ estimates.

\subsection{Classical vs quantum BHs: (the importance of) Bayes factors}

A discussion of the Bayes factors is pivotal when interpreting the inferred posteriors for the parameters. 
A by-product of a full Bayesian analysis is the relative Bayes factor between the hypotheses 
that i) a signal corresponding to the chosen model is present in the data and ii) the data are described by Gaussian noise. 
In the notation of Sec.~\ref{subsec:Bayesian data analysis}, we may consider $\mathcal{H}_i \equiv GR$, $\mathcal{H}_j \equiv nGR_F$, and $\mathcal{H}_k \equiv nGR_A$: we can test the FK model of QBH against the classical GR model of BH using the Bayes factors produced by each analysis.
Given a choice of common priors for the parameters, we can test a pair of models against each other by looking at the natural logarithm of their Bayes factor.
The ratio of their signal-to-noise log Bayes factor logB$^s_n$, reported in Table~\ref{tab:logBF}, yields the log Bayes factor of the two models which, assuming equal model priors, is equal to their odds' ratio.
We can adopt the \emph{Jeffreys' scale} (see Appendix B of \cite{JEFFREYS}) to interpret the obtained odds' ratio values.

When analyzing GW150914 assuming the $GR$ model and uninformative priors for $M_f$ and $a_f$, we obtain logB$^{GR}_{noise} = 46.8$~\footnote{Note that this number differs from the one reported in~\cite{CARULLO-DELPOZZO-VEITCH}, 
since here we adopt a likelihood function that is based on the truncation procedure outlined in~\cite{Isi:2019aib}.}.
Considering the same set of priors, we can compare it against the Bayes factor obtained from the analysis of GW150914 assuming either $nGR_F$ (logB$^{nGR_F}_{noise} = 44.3$) or $nGR_A$ (logB$^{nGR_A}_{noise} = 46.7$).
Firstly, we note a 2.4 e-fold preference of $nGR_A$ over $nGR_F$:  this may be explained with the latter model incurring an \emph{Occam factor penalty} (see also Ref.~\cite{CalderonBustillo:2020tjf} for a similar discussion on GW150914 ringdown tests), since an increased prior volume generally results in a decreased prior density around the maximum likelihood region, yielding a lower evidence. 
Also, given the very similar posteriors of $\alpha$ obtained with these two models, we may interpret these Bayes factors as a corroboration of the fact that Eq.~(\ref{eqn:quantum_QNM_fundamental_tau}) is a fairly good working hypothesis. Secondly, the $GR$ and $nGR_A$ log Bayes factors overlap within their errors (all the single-event log Bayes factors throughout this paper have an uncertainty of 0.1, as discussed in Sec.~\ref{sec:GWTC section}-\ref{sec:Population section}), implying that we cannot draw any conclusion from the analysis of GW150914 on which model is preferred.

Employing the more restrictive IMR priors, a $GR$ analysis yields logB$^{GR}_{noise} = 50.9$, while $nGR_F$ and $nGR_A$ models have logB$^{nGR_F}_{noise} = 46.9$ and logB$^{nGR_A}_{noise} = 49.5$, respectively.
Now the $GR$ logB has raised by 4.1 e-folds compared to the uninformative-prior case, while the alternative models have gained 2.6 and 2.8 e-folds: the increase common to all models is expected as more informative priors have been chosen, and so the prior density is higher around the region of high likelihood.
Overall, the difference between $GR$ and $nGR_A$ is now increased up to 1 e-fold, showing a very feeble preference towards GR.

In conclusion, GW150914 data are not sufficient by themselves to constrain $\alpha$, nor to discriminate between GR and alternative hypotheses. 
Signals with SNRs that will be detected by the Advanced interferometers at increased sensitivity, or by future interferometers \cite{ET, LISA}, may be loud enough that in principle  they could both validate the FK model \emph{and} constrain the value of $\alpha$, or disprove it altogether.
As we will show in Sec.~\ref{sec:Population section}, without a massively loud signal an alternative route is to coherently consider the information encoded in different independent GW events. 
Exploiting the fact that $\alpha$ is a universal constant, measurements coming from multiple events -- even if relatively weak -- can be combined to obtain a \emph{joint} posterior PDF. 

Since the uninformative-prior choice fully accounts for the correlations between $\alpha$ and the BH parameters and it does not assume a classical pre-ringdown emission (thus still allowing for potential inspiral-only imprints of the area quantisation, like tidal heating~\cite{CARDOSO-FOIT-KLEBAN, AGULLO2020}), we deem this more conservative choice preferable and employ it in the remainder of the study. 

Finally, we assessed with the loudest ringdown event that the measurement of $\alpha$ is not affected by the assumption of the ansatz for $\tau_0$. For this reason, along with the mild logB preference for $nGR_A$ over $nGR_F$, in the remainder of this work we assume $nGR_A$ as benchmark for our $GR$-alternative model, thus reducing the computational cost of the analysis. Since from now on we will only be considering the $nGR_A$ model as our competing model to $GR$, we relabel $nGR_A \equiv nGR$.

\section{Inferring $\alpha$ from O1-O2-O3a events}\label{sec:GWTC section}

\begin{figure*}[!tb]
\includegraphics[width=0.5\textwidth]{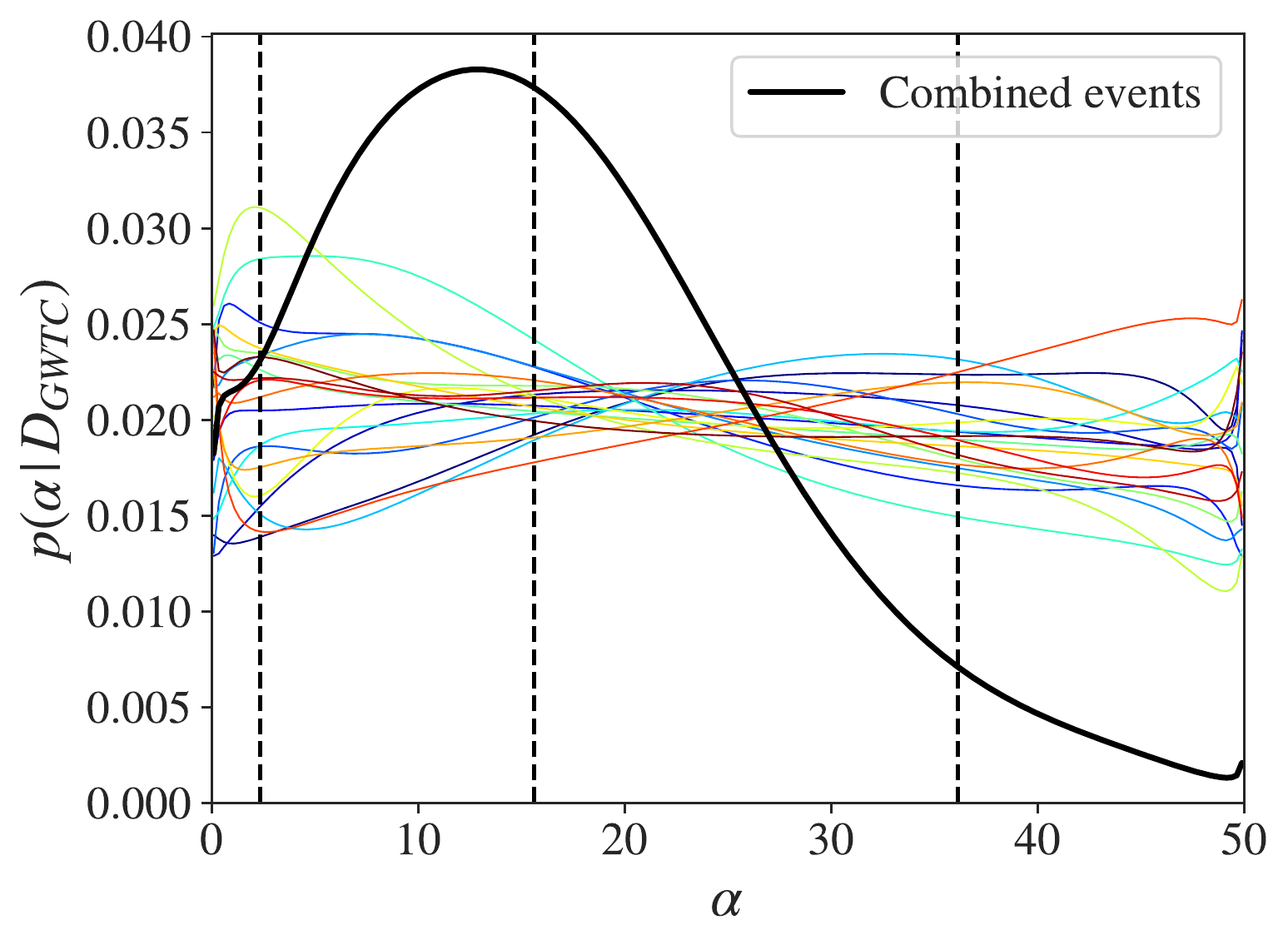}
\includegraphics[width=0.5\textwidth]{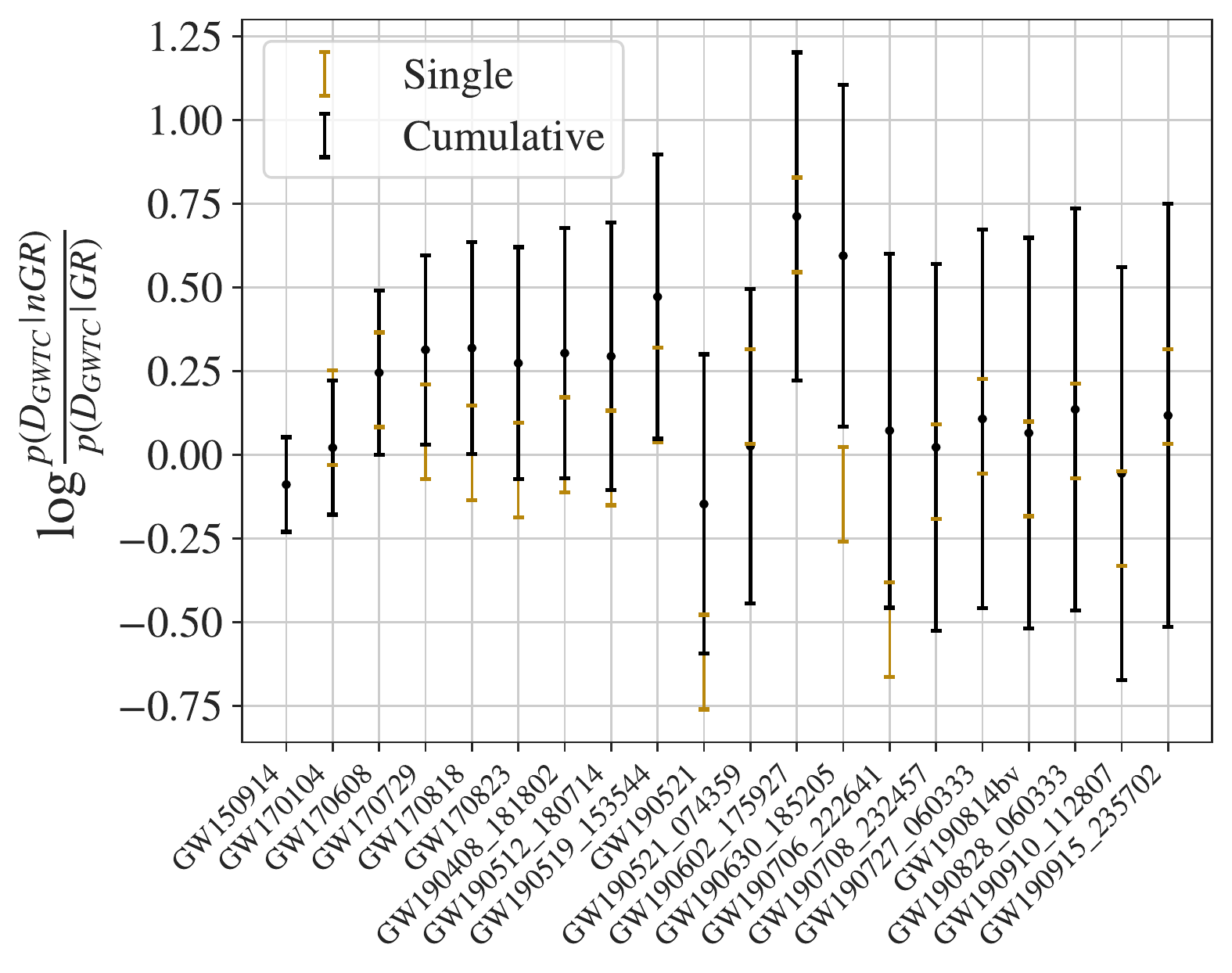}
\caption{Left panel: single-event posterior distributions (coloured thin lines) for the $\alpha$ parameter and combined-event posterior (black thick line) obtained with a Dirichlet Process Gaussian-mixture model using GWTC ringdown data $D_{GWTC}$. Dotted vertical lines correspond to the median and $90\%$ CRs. The measured value is $\alpha = 15.6^{+20.5}_{-13.3}$. The lower region of the parameter space is preferred by the combined posterior.
Right panel: single (gold bar) and cumulative (black bar) logarithm of the Bayes factors for each of the LVC ringdown detections. The uncertainty on each single-event Bayes factor is 0.1; the uncertainty on the final cumulative Bayes factor is $\pm 0.6$. No preference between the two competing models is observed as the final cumulative Bayes factor is $0.1 \pm 0.6$.}
\label{fig:alpha_GWTC_events} 
\end{figure*}

\begin{table}
\caption{Summary of the signal-to-noise Bayes factors for the $GR$ and $nGR$ models obtained with uninformative priors for the final mass and spin. The events included in the analyses must have informative posteriors and a positive logB$^{GR}_n$. The numerical statistical error on the logB is $\pm \, 0.1$.}
\begin{indented}
\item[] \begin{tabular}{@{}lcc}
\hline\hline
\multicolumn{3}{c}{O1O2O3a ringdown events} \\
\hline
\hline
\multicolumn{3}{c}{Uninformative priors} \\
\hline
Event & logB$^{GR}_n$  & logB$^{nGR}_n$\\
\hline
GW150914   & $46.8$ &  $46.7$\\
GW170104     & $9.3$  &  $9.4$\\
GW170608     & $4.8$  &  $5.0$\\
GW170729 	 & $1.0$  &  $1.1$\\
GW170818   & $2.7$ &    $2.7$\\
GW170823     & $21.1$  &  $21.1$\\
GW190408\_181802   & $19.6$  &  $19.6$\\
GW190512\_180714   & $11.1$  &  $11.2$\\
GW190519\_153544 	 & $45.9$  &  $46.0$\\
GW190521                  & $56.5$ &  $ 57.0$\\
GW190521\_074359    & $34.9$ &  $35.1$\\
GW190602\_175927   & $18.9$  &  $19.5$\\
GW190630\_185205   & $50.4$  &  $50.3$\\
GW190706\_222641   & $21.5$  &  $21.0$\\
GW190708\_232457    & $5.9$  &  $5.9$\\
GW190727\_060333   & $22.6$  &  $22.7$\\
GW190814                  & $8.0$  &  $7.9$\\
GW190828\_060333   & $9.9$  &  $9.9$\\
GW190910\_112807   & $24.4$  &  $24.2$\\
GW190915\_235702   & $24.9$  &  $25.1$\\
\hline\hline
\end{tabular}
\end{indented}
\label{tab:O1O2O3_events}
\end{table}

We now investigate the possibility of testing the FK model combining the information from the GW BBH events reported in the Gravitational-Wave Transient Catalogs GWTC-1~\cite{GW-O1-O2-catalog} and GWTC-2~\cite{O3a-CBC-CATALOG}, using the GW strain data publicly available at the Gravitational Wave Open Science Center~\cite{OPEN-DATA-GWTC1, GWOSC}. The events that we analyze are reported in Table~\ref{tab:O1O2O3_events}, along with the logBs associated to the $GR$ and $nGR$ models. This list differs in some events from that reported in~\cite{O3a-TGR-PAPER} because here we adopt a different GR template to classify the presence of a ringdown signal in the data: in this work we consider only the GW events for which an analysis with our $GR$ model shows an informative posterior \emph{and} a positive signal-to-noise log Bayes factor. We analyze each event in a similar way as done for GW150914. We will then combine the posteriors of $\alpha$ from multiple events taking advantage of a Dirichlet Process Gaussian-mixture model, a fully Bayesian non-parametric method, to build PDF out of single-event samples (for a pedagogical introduction, see~\cite{DELPOZZO2018}). The obtained posterior will be assessed by the cumulative odds' ratio of the two competing models.

In the left panel of Fig.~\ref{fig:alpha_GWTC_events} we show both single-event posteriors (coloured thin lines) and the associated combined PDF for $\alpha$ (black thick line). 
As already seen with GW150914, single-event posteriors provide only little information. Indeed we observe that the posteriors for the final mass are only broadened compared to the ones in~\cite{O3a-TGR-PAPER}, as expected by the increase of the dimension of the parameter space, while the posteriors for the final spins are essentially identical due to the mild dependence of the QNM parameters on the latter.
However, the Dirichlet-process combined posterior confidently excludes the upper portion of the prior range, constraining the area quantisation parameter to be $\alpha = 15.6^{+20.5}_{-13.3}$. 
On the right panel we plot the logarithm of single-event Bayes factors (gold bar) together with the \emph{cumulative} Bayes factor (black bar) obtained combining each previous (independent) event. 
Assuming equal prior probability to each model, so that $\frac{p(nGR| I)}{p(GR| I)} = 1$, the plotted quantities are the single and cumulative odds' ratios $O^{nGR}_{GR}$ (recall Eq.~(\ref{eqn:odds_ratio})). 
A conservative estimate of the numerical uncertainty of the single Bayes factors has been assigned repeating with 10 different seeds the analysis of the event with lowest logB, thus taking their standard deviation, yielding a estimate of $\pm 0.1$. Propagating the single event uncertainty to the overall analysis, we conclude that the cumulative Bayes factor is $0.1 \pm 0.6$: the considered events are not significant enough to allow us to discriminate between GR and the FK model.

It should be stressed that if considering only the combined $\alpha$-posterior, without taking into account the cumulative Bayes factor, one may be tempted to conclude that the events are providing a measurement of the area quantisation parameter, implicitly implying that the FK model is preferred by the data (because if GR were correct one would not expect to recover a peaked distribution for $\alpha$). 
This reasoning is incorrect, because marginalised posteriors on a specific quantity are not a measure -- not even a relative one -- of the goodness of fit in a Bayesian framework, unlike the Bayes factor. 
This statement will become apparent in the next section, where we show how a behaviour similar to what is observed with interferometric data can be reproduced by a population of \emph{classical} BHs, that is, even when GR is the underlying theory describing the gravitational radiation emission.

\section{Testing the BM conjecture with a simulated BBH population}\label{sec:Population section}

Having obtained uninformative results from current observed events, we turn to a systematic study employing simulated signals, or ``injections", coming from an astrophysical population of GW150914-like events detected by the LIGO-Virgo network at their design sensitivity.
This study will be instrumental not only to test the feasibility of our method and understand if the conjecture can be tested with current GW detectors, but also to better interpret the results obtained with real data and classify the features that can show up when performing a population analysis. 
As it will be apparent, drawing meaningful conclusions from a population analysis requires to look at global figures of merit (like Bayes factors) and not only to marginalised one-dimensional distributions of the parameters of interest. 

Being a proof-of-concept study, we realise zero-noise signals to make our analysis independent of a specific noise realization, using the estimated design-sensitivity power spectral density to compute our likelihood function, Eq.~(\ref{eqn:likelihood_function}).

\subsection{Generation of population parameters} 

To create a population of GW150914-like remnant BHs we generate a set of random parameters from uniform distributions as follows: $\alpha' \in [0,2\pi]$, $\cos \delta' \in [-1,+1]$, $\psi \in [0,\pi]$, $\cos \iota \in [0.5,1]$, $\log D_L \in [\log 300, \log 1000]$, $\phi \in [0,2\pi]$, $a_{1,2} \in [-1,+1]$, $M_{1,2} \in [25,50] M_{\odot}$. 
Here $a_{1,2}$ and $M_{1,2}$ are spins and masses of the progenitor BHs. 
We use the BBH final-state fitting formulas presented in~\cite{JIMENEZ-FORTEZA-KEITEL-HUSA-HANNAM-KHAN-PUERRER} to generate the remnant $M_f$ and $a_f$. 
To obtain signals with a realistic ringdown loudness, we employ the numerical relativity (NR) fits of~\cite{LONDON2018} -- describing remnant BHs generated by the collision of non-precessing BBH -- to generate the amplitudes $\mathcal{A}(\eta, \chi_s)$ as a function of the symmetric mass ratio $\eta = (M_1  M_2)/(M_1 + M_2)^2$ and the symmetric spin combination $\chi_{s} = (M_1 a_1 + M_2 a_2)/(M_1 + M_2)$ of the progenitor BHs. 
Finally, for simplicity and to consistently apply the assumed non-precessing symmetry of the amplitudes, we will set $\tilde{\mathcal{A}}' = \tilde{\mathcal{A^*}}$~\cite{BLANCHET2006}.
Thus, an event $D_i$ is generated with a set of 8 parameters $\vec{\theta} = \{\alpha', \delta', \psi, D_L, \mathcal{A}, \phi, M_f, a_f \}$, which are then used in the waveform model Eq.~(\ref{eqn:Kerr_template}) to generate a simulated strain $h(t)$, so that our data $d(t)$ is composed of zeros plus the mock signal.

A given set of parameters $\vec{\theta}$ can be used to generate either a \emph{GR} or a \emph{nGR} waveform. In case of a \emph{nGR} injection, the set of injected parameters includes one more parameter, $\alpha$, which must be fixed to some value. By the same token, we can recover a given signal by assuming a \emph{GR} or a \emph{nGR} template, in the latter case allowing for a uniform $\alpha$-prior, $\alpha \in [0,50]$. 
The SNR of the injected signals goes roughly from $10$ to $95$ and will be specified for each population. Finally, the uncertainty on the single-event logB is evaluated choosing a random injection and adopting the procedure described in Sec.~\ref{sec:GWTC section}, yielding a value of $\pm0.1$.

\subsection{Prior choice for simulated events}

When analyzing simulated events, attention must be paid to the choice of the priors to be adopted in the analysis: to avoid biases in a population analyses, the priors must match the same distribution from which the parameters $\vec{\theta}$ are generated. 
A choice of uniform priors for $M_f$, $a_f$, and $\mathcal{A}$ in our case would not be an optimal choice since, when simulating a BH population distribution, it is appropriate to use progenitor parameters. 
Realistic distributions on remnant parameters are then generated using NR relations taking as input initial masses and spins drawn from the progenitor distributions. This operation does not produce uniform distributions on the remnant parameters, thus the corresponding priors need to be set accordingly.
Without accounting for this additional prior weight, we verified the presence of the aforementioned biases in the estimation of remnant parameters.

\subsection{Population of GR BHs}

\begin{figure*}[!tb]
\includegraphics[width=0.5\textwidth]{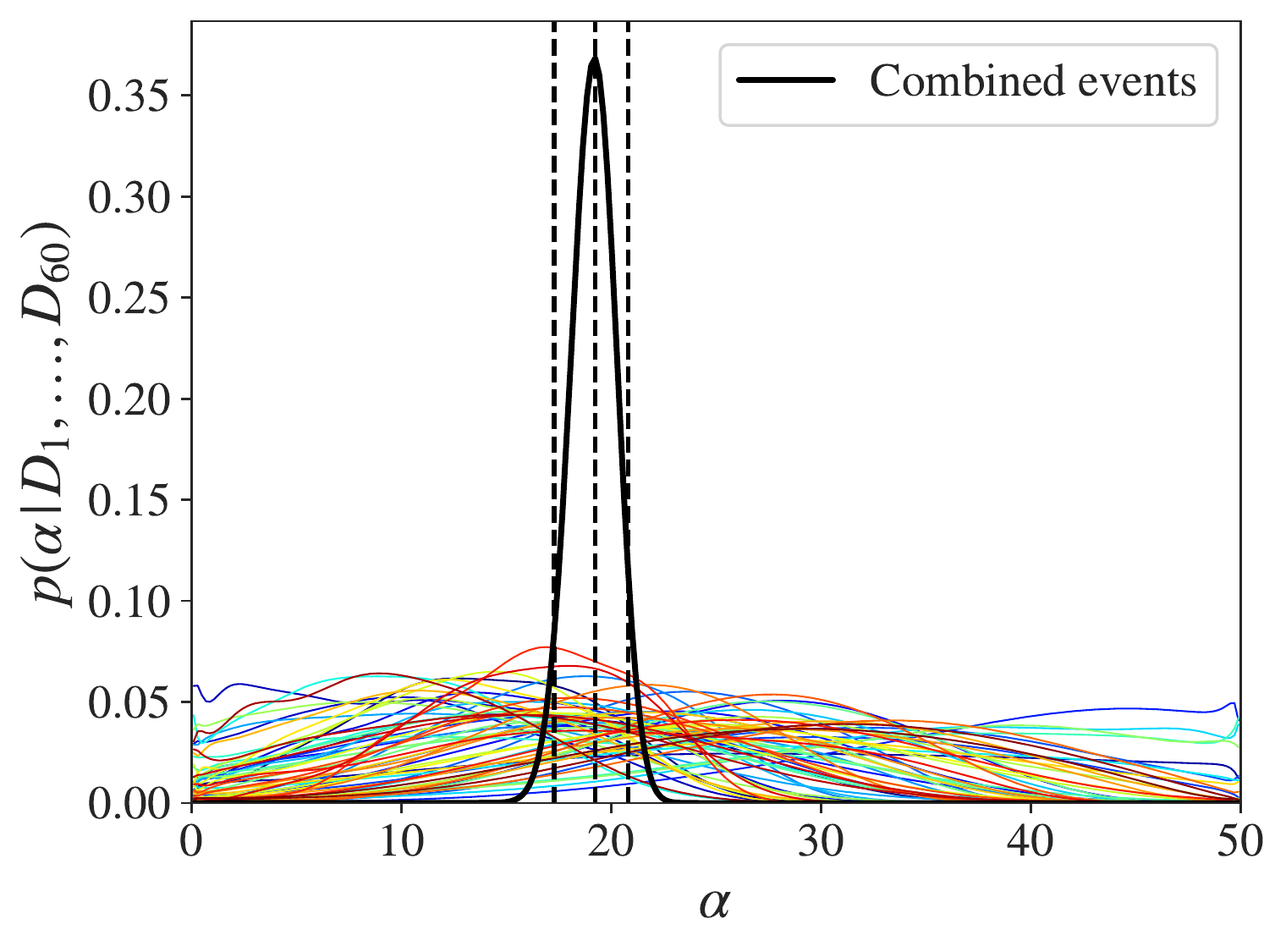}
\includegraphics[width=0.5\textwidth]{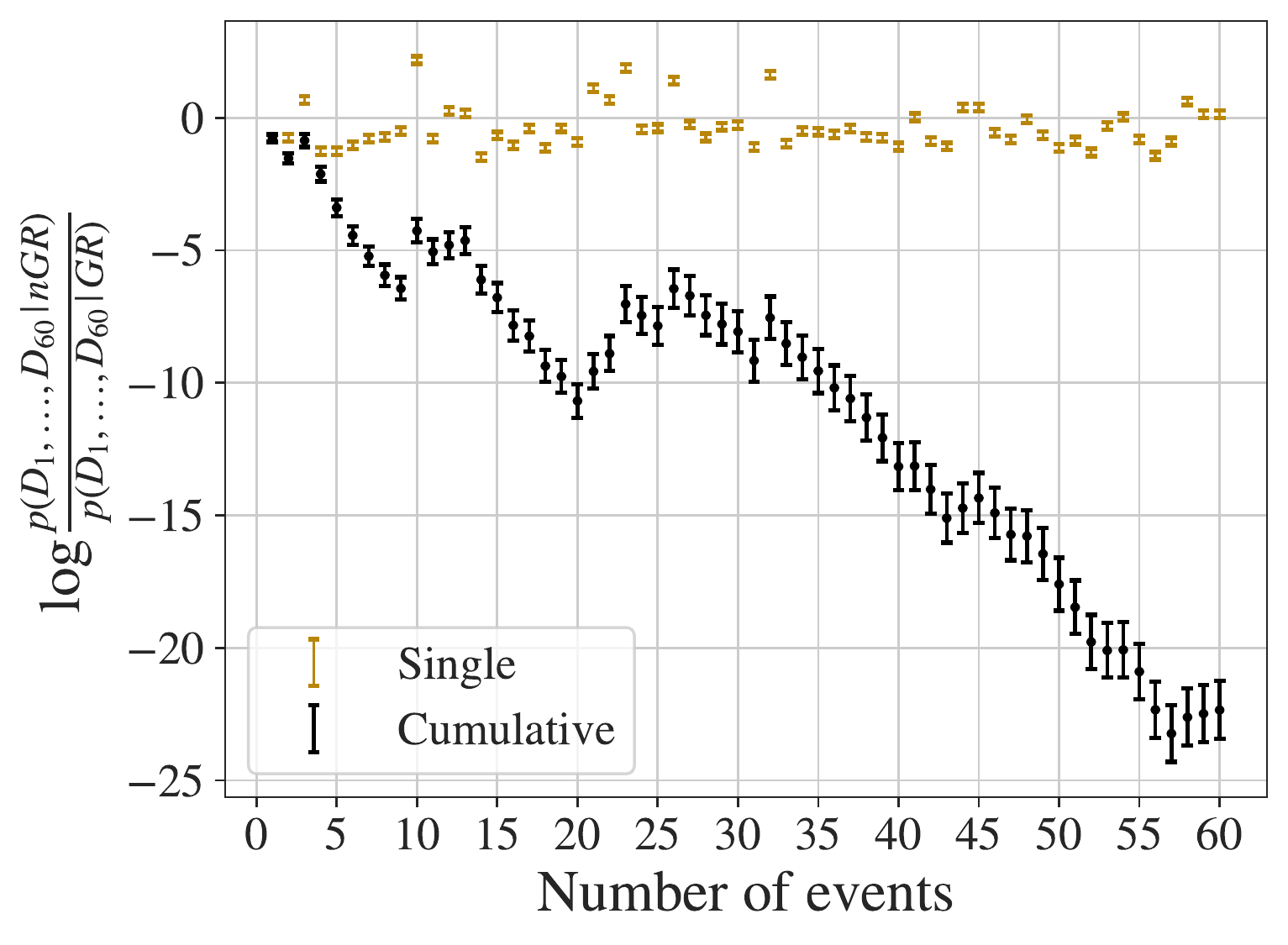}
\includegraphics[width=0.5\textwidth]{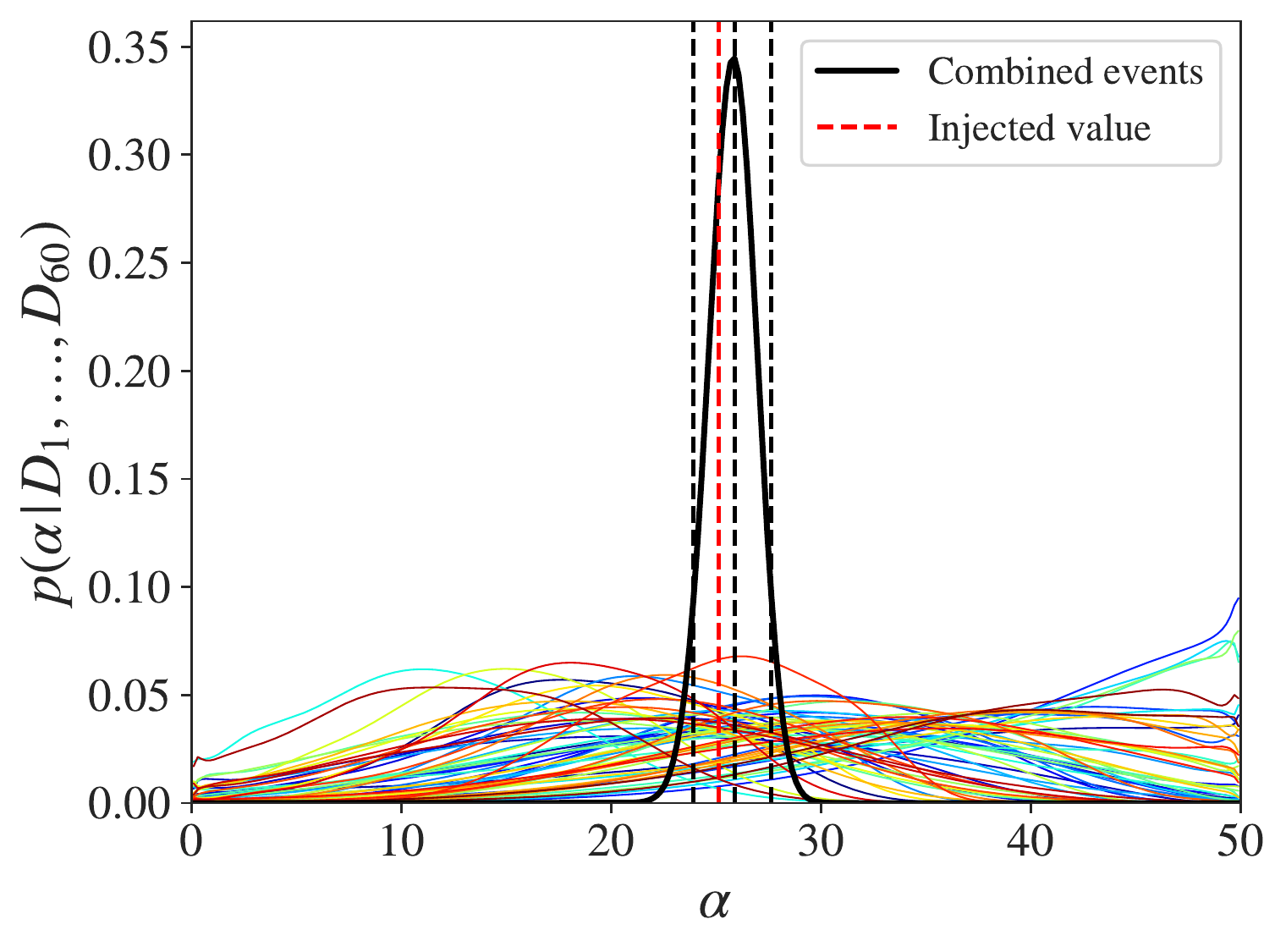}
\includegraphics[width=0.5\textwidth]{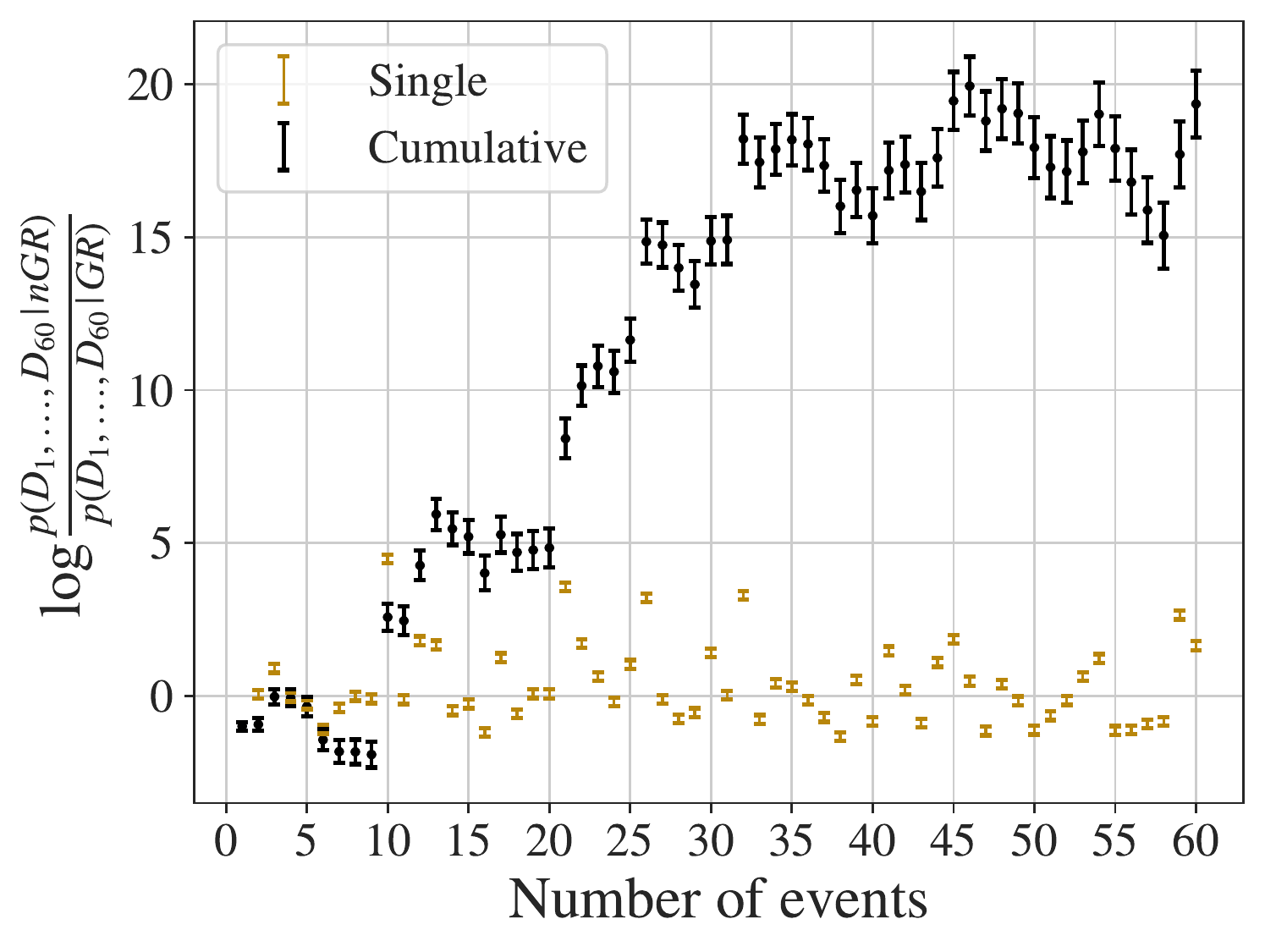}
\caption{Upper left panel: joint PDF for $\alpha$ (black thick curve) for a population of 60 \emph{GR} BHs recovered using a \emph{nGR} template, obtained by combining the single PDFs (colored thin curves). Dotted lines indicate median and 90\% CR values: $\alpha = 19.0^{+2.0}_{-2.0}$, which would suggest a measurement of the parameter even if no value of $\alpha$ was injected.
Upper right panel: cumulative Bayes factors of $nGR$ over $GR$ for a population of 60 \emph{GR} BHs. Gold bars correspond to single-event Bayes Factors, while black bars to cumulative Bayes factors as the data set is extended to include each additional event. The uncertainty on each single event Bayes Factor is $\pm 0.1$; the uncertainty on the final cumulative Bayes factor is $\pm 1.1$. $GR$ is strongly favored: the Bayes factors allow to correctly identify the injected model.
Lower left panel: joint PDF for $\alpha$ (black thick curve) for a population of 60 \emph{nGR} QBHs recovered using a \emph{nGR} template, obtained by combining the single PDFs (colored thin curves). The recovered value is: $\alpha = 25.7^{+2.0}_{-1.8}$, peaking around the injected value (red dotted line): $\alpha \simeq 25.1$.
Lower right panel: cumulative Bayes factors of $nGR$ over $GR$ for a population of 60 \emph{nGR} QBHs. \emph{GR} is highly disfavored, as expected in this case.}
\label{fig:joint_alpha_cumulative_logB_injections} 
\end{figure*}

\begin{figure}[h]
\center
\includegraphics[width=0.7\textwidth]{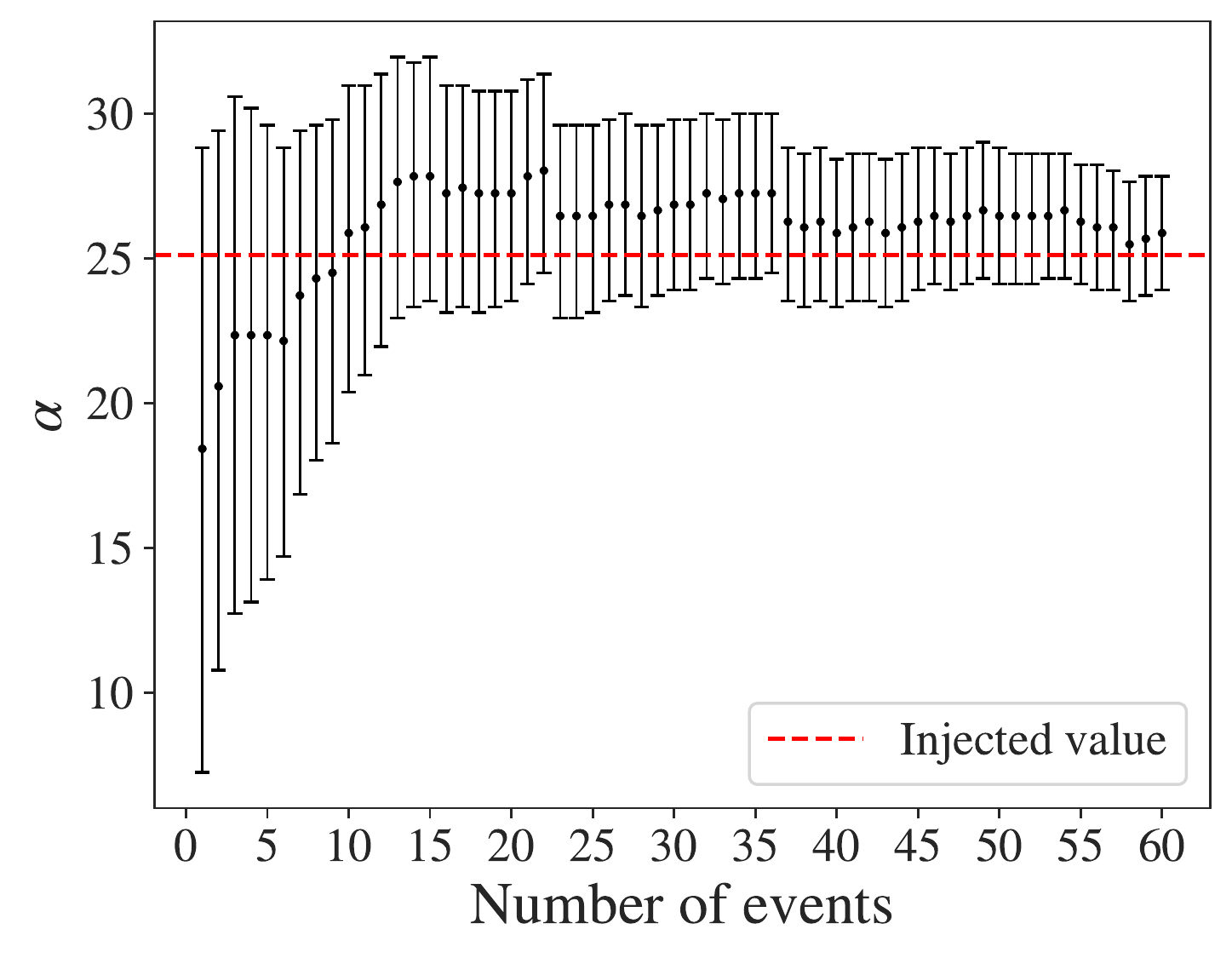}
\caption{90\% CRs and median value of the joint PDF for $\alpha$ as different $nGR$ events are analysed and combined.}
\label{fig:alpha_nGR_rec_posterior_vs_events} 
\end{figure}

We first simulate a population of 60 GR BHs emitting $GR$ ringdown signals that we recover in two alternative ways: i) assuming that $GR$ is correct; ii) assuming that $nGR$ is correct. 
The strategy is to analyse each event adopting two mutually exclusive waveform models and see which one best reconstructs the injected waveform. 
Since we are injecting classical signals, we expect \emph{GR} to be preferred by the analyses. 
According to what we learned in Secs.~\ref{sec:GW150914 section} and \ref{sec:GWTC section}, we do not expect to get this information from single-event analyses, but instead from the combination of them.

This happens to be the case, as shown in the upper right panel of Fig. \ref{fig:joint_alpha_cumulative_logB_injections}, where the logarithm of the ratio of Bayes factors of each event (gold bars) is reported together with the cumulative Bayes factor (black bars), under the assumption that the events are independent. 
Assuming equal priors for the models, we can interpret the plotted quantities as odds' ratios:  20 events are enough to make the \emph{GR} model preferred $\simeq e^{10.7\pm0.6}$ times more than the \emph{nGR} model.
This difference is even increased when the total population of 60 events is considered: \emph{GR} is preferred almost $\simeq e^{22.3\pm1.1}$ times more than the \emph{nGR} model. Regarding the strength of the injections, 47\% (77\%) of the 60 events have SNR $\leq 30~(50)$, with lowest (highest) SNR equal to 11.3 (95.9).

A relevant point of discussion is put forward by the combined posterior shown in the upper left panel of Fig. \ref{fig:joint_alpha_cumulative_logB_injections}: the black curve is the joint PDF for $\alpha$ produced by the analyses where \emph{nGR} is assumed to be correct. It shows a strongly peaked posterior distribution with $\alpha = 19.2^{+1.6}_{-1.9}$; 
however, having simulated \emph{GR} ringdown signals, there is no underlying parameter $\alpha$ corresponding to this population. This result could naively raise concerns on the analysis at first, but on a closer look there is no inconsistency with it. If we assume that $\alpha$ should be there, even though not present in the data, Bayesian inference will try to infer the value which best matches the data that are being analysed.
Furthermore, as already stated in Sec.~\ref{sec:GWTC section}, marginalised posteriors by themselves do \emph{not} legitimate the goodness of a Bayesian parameter estimation, as they give only a partial analysis: Bayes factors should always also be looked at, when available. 
Indeed, the cumulative Bayes factor indicates that \emph{GR} is extremely preferred over \emph{nGR} in describing the 60 injected events. This invalidates the conclusions inferred assuming the \emph{nGR} model, suggesting not to trust the combined posterior in the upper left panel of Fig.~\ref{fig:joint_alpha_cumulative_logB_injections}.
This result helps us interpret the combined analysis of LVC detections reported in Sec.~\ref{sec:GWTC section}. It shows that even if GR is correct, a posterior similar to the one plotted in Fig.~\ref{fig:alpha_GWTC_events} is to be fully expected. 
The discriminator between competing models is the Bayes factor, which in the case of real events -- unlike the simulations just discussed -- does not allow yet to discriminate among the two competing models.

\subsection{Population of Bekenstein QBHs}

We now explore the complementary case: a simulation of 60 \emph{nGR} signals generated by QBHs with a fixed value of $\alpha$. We fix $\alpha =8 \pi$ \cite{BEKENSTEIN1973, MAGGIORE2008, MEDVED, VAGENAS, AGULLO2020}, although our qualitative conclusions are independent of the specific figure chosen to create the simulated population. 
This case helps us to determine whether such a model can be distinguished from $GR$ with the current detector infrastructure -- in principle one could conceive of the possibility of a $nGR$ model for which a $GR$ template always produces a good fit of the signal at low signal-to-noise (SNR) -- and how many detections will be needed to confidently claim its validity. 
Following the same procedure outlined before, we now inject \emph{nGR} ringdown signals and subsequently analyse the data assuming either \emph{GR} or \emph{nGR}. 
What we would like to know is: i) whether or not the \emph{nGR} template is preferred over the \emph{GR} one (as it should); ii) whether the injected $\alpha$ is correctly recovered in the parameter estimation. 
The answer to these questions is positive in both cases: as shown in the lower right panel of Fig. \ref{fig:joint_alpha_cumulative_logB_injections}, the \emph{nGR} model is strongly supported by the cumulative odds' ratio, which after 60 events is $O^{nGR}_{GR} \simeq e^{19.4\pm1.1}$. 
The joint PDF posterior is reported in the lower left panel of Fig. \ref{fig:joint_alpha_cumulative_logB_injections}, with $\alpha = 25.9^{+1.7}_{-2.0}$, consistent with the injected value $\alpha \simeq 25.1$.
The injected signals have an SNR which in 50\% (83\%) of the cases is $\leq$ 30 (50), the faintest being 10.7 and the loudest being 90.2.
Fig.~\ref{fig:alpha_nGR_rec_posterior_vs_events} shows the shrinkage of the $90\%$ CRs of the joint posterior as we increase the number of analysed sources, with the median value correctly converging towards the injected one.

\subsection{Stealth biases}

At this point, a natural question to ask is whether we can quantify the \textit{stealth biases} ~\cite{YUNES-PRETORIUS, VALLISNERI-YUNES, VITALE-DEL_POZZO} caused by using a $GR$ model when performing parameter estimation on $nGR$ data. In other words, if the heuristic model of FK correctly described the ringdown signals that we will observe, would such deviations from GR predictions be flagged by standard GR tests \cite{TGR-LVC2016, TGR-O1-O2} performed by the LVC?

\begin{figure}[h]
\center
\includegraphics[width=0.7\textwidth]{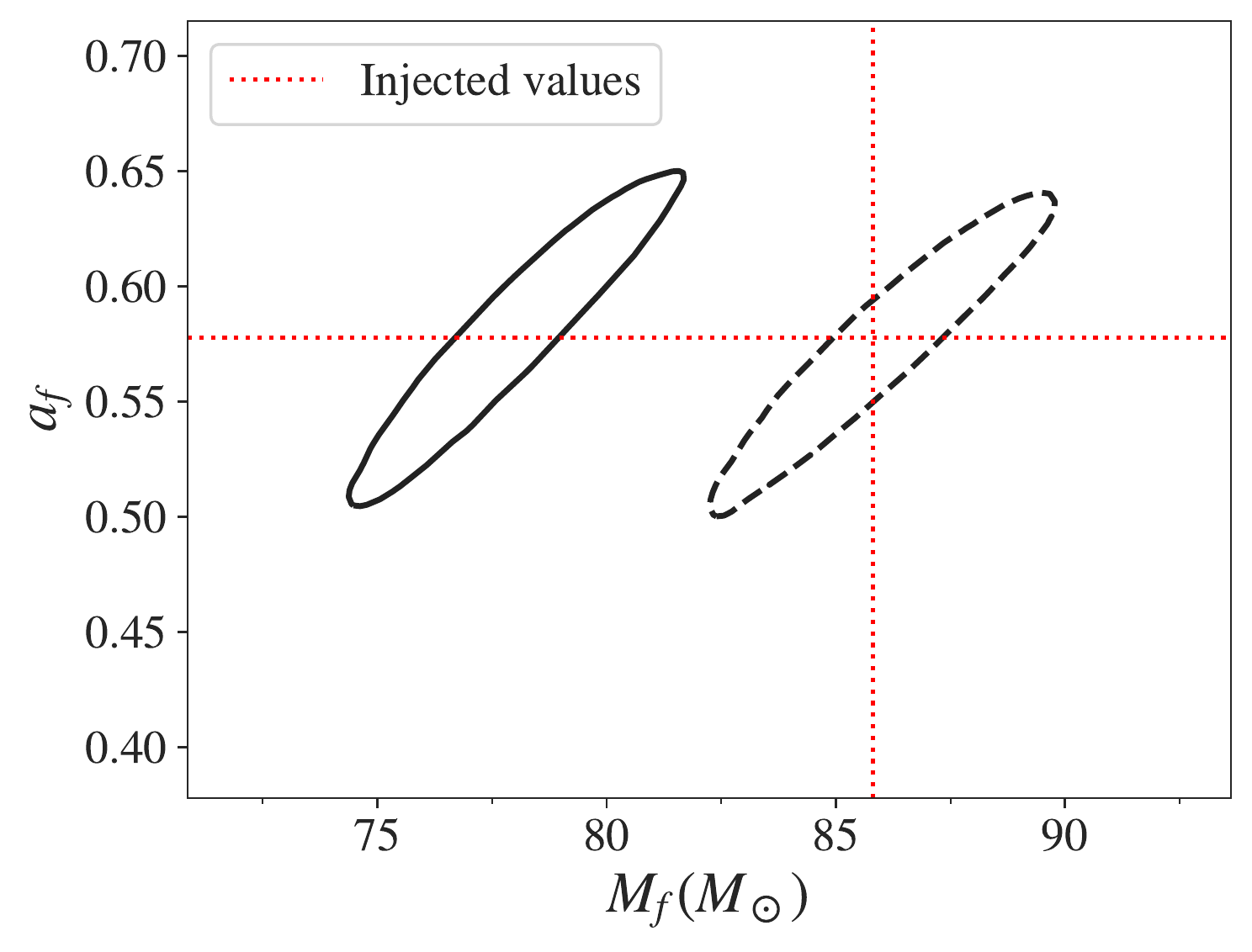}
\caption{Two-dimensional $90\%$ CR posterior distributions of the BH intrinsic parameters assuming  $GR$ template for two different injections: i) when a \emph{GR} signal is injected, the posterior (dashed contour) correctly includes the injected values (dotted lines); ii) when a $nGR$ signal with identical parameters but $\alpha = 8 \pi$ is injected, the algorithm tries to fit the signal with the incompatible model, and consequently the posterior (solid contour) shows a visible bias in the final mass. The recovered SNRs are 69 and 65, respectively. This kind of bias would naturally show up in an IMR consistency test.}
\label{fig:stealth_bias} 
\end{figure}

A first specific example of stealth bias is shown in Fig.~\ref{fig:stealth_bias}, where we report the measurement of final mass and spin from two simulated ringdown signals taken from the $GR$ and $nGR$ populations previously generated. 
These two signals have been generated using the same parameters (red dotted lines) and are identical except for the area quantisation assumption (with $\alpha = 8 \pi$), while the recovery is always performed assuming that $GR$ is the correct model. 
While the analysis of the \emph{GR} injection (black dashed contour) correctly recovers the injected values of mass and spin, the same analysis on the \emph{nGR} signal (black solid contour) shows a shifted posterior in $M_f$, with the injected values well outside its $90\%$ CRs. 
This shift arises because of the correlation of $\alpha$ with the QBH mass and spin: the $GR$ template is trying to account for the presence of the additional parameter by adjusting the BH intrinsic parameters, biasing the mass with respect to the injected value.

\begin{figure*}[!tb]
\includegraphics[width=0.51\textwidth]{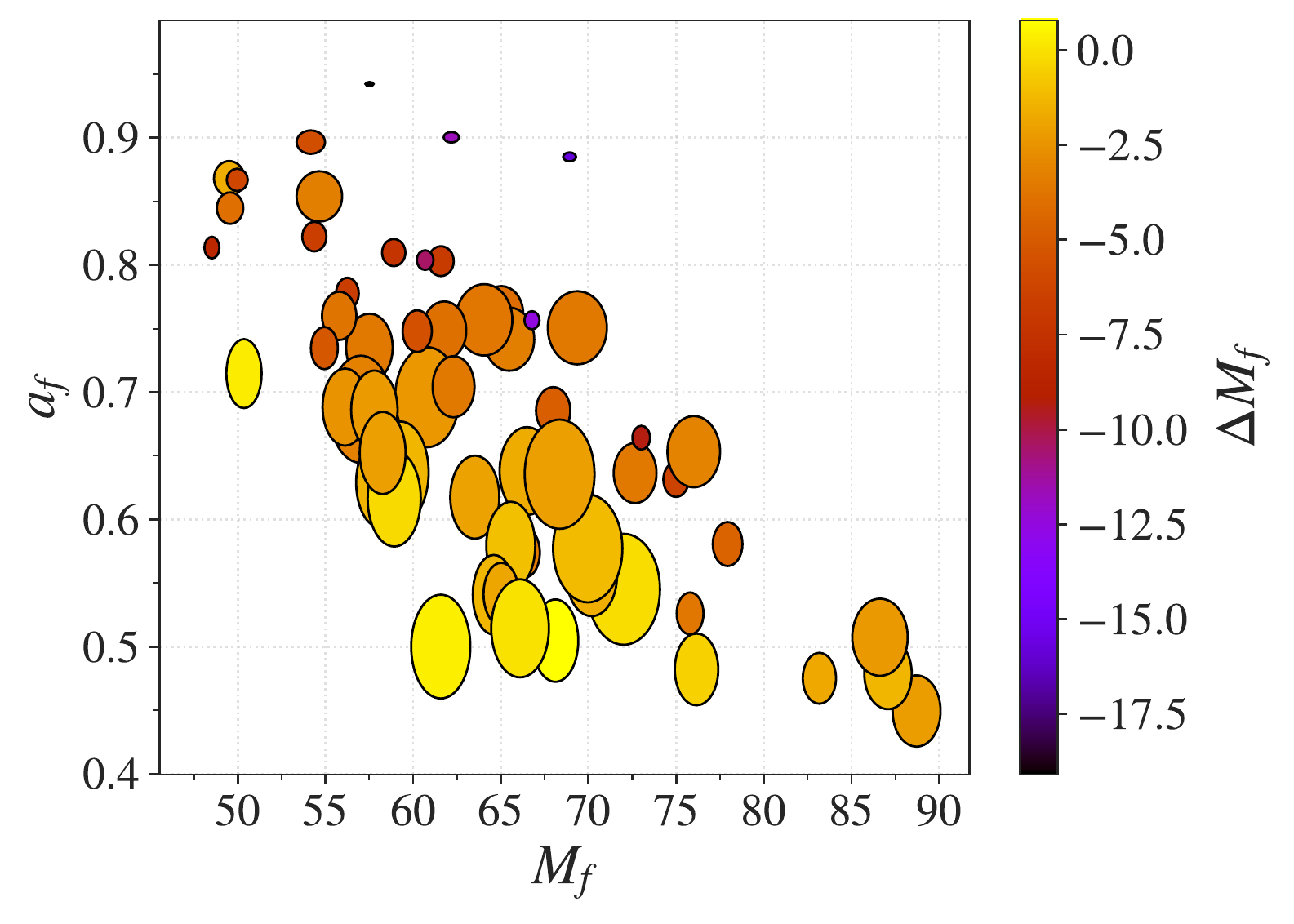}
\includegraphics[width=0.49\textwidth]{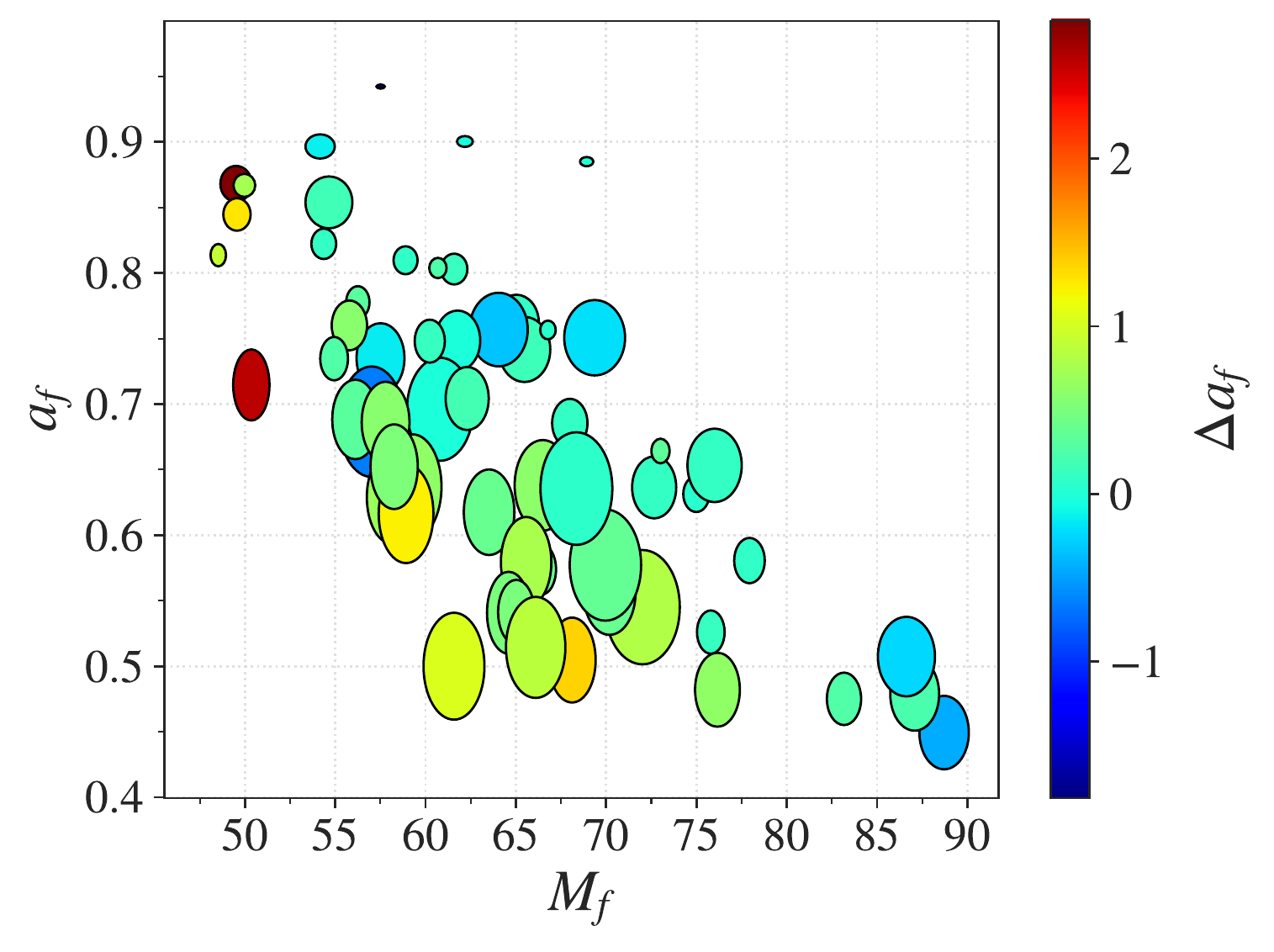}
\caption{Left panel: effect size $\Delta M_f = (\overline{M}_f - M_f^{inj})/\sigma_{M_f}$as function of the injected values of $M_f$ and $a_f$. 
Each ellipse represents the values of the single-event reconstructed mass and spin, centered on median values and with axes proportional to their standard deviation, while the color scales from the smallest to the largest value of the effect size according to the scale reported on the vertical bar. We see dominantly a negative bias on the measure of the mass.
Right panel: the same as in the left, but now for the effect size $\Delta a_f = (\overline{a}_f - a_f^{inj})/\sigma_{a_f}$. In this case no predominant bias is observed.
In both panels all the $nGR$ signals are recovered assuming that GR is correct.}
\label{fig:bias_mass_spin_NGR} 
\end{figure*}

To systematically quantify the fraction of cases in which a significant bias could be observed and identify the region of the parameter space where this effect is more prominent, for each event we can define the ``effect size'' \cite{VITALE-DEL_POZZO} of a generic parameter $p$ as
\begin{equation}\label{eqn:effect_size_bias}
\Delta p = (\overline{p} - p^{inj})/\sigma_{p} \,,
\end{equation}
where $\overline{p}$ is the median value of the reconstructed posterior, $p^{inj}$ is the injected value, and $\sigma_{p}$ is the standard deviation evaluated from the reconstructed posterior samples of the parameter. 
Large values of $\Delta p$ encode large deviations from the true value.
In the left panel of Fig.~\ref{fig:bias_mass_spin_NGR} we show the effect size $\Delta M_f = (\overline{M}_f - M_f^{inj})/\sigma_{M_f}$ as a function of the $M_f$ and $a_f$ parameter space, evaluated on the previously analysed \emph{nGR} population when a \emph{GR} model is assumed. 
We find that 70\% of the \emph{nGR} injected final masses present a bias larger than $2 \sigma_{M_f}$ ($|\Delta M_f| \geq 2$). The sign of the bias is dominantly negative (95\% of the events): the effect of assuming $GR$ is to decrease the value of $M_f$, as evident from the example showed in Fig.~\ref{fig:stealth_bias}. 
In the right panel of Fig.~\ref{fig:bias_mass_spin_NGR} we show the same quantity for the final spin, $\Delta a_f = (\overline{a}_f - a_f^{inj})/\sigma_{a_f}$, which suffers much less from a stealth bias: just 2 events out of 60 show a (positive) effect size greater than $2$.

An example of the fact that a bias is likely to affect more $M_f$ than $a_f$ has already been given in Fig.~\ref{fig:stealth_bias}, where the $nGR$ posterior is essentially only shifted along the $M_f$ axis compared to the $GR$ one. This is due to the stronger correlation among the $M_f$ and $\alpha$ parameters, as well as to the better precision in measuring $M_f$ with respect to $a_f$.

For a given mass and spin the Bekenstein QNM frequency is higher compared to that of the GR fundamental frequency, resulting in lower recovered masses (95\% of the cases).  
The same reasoning can be applied to $a_f$, since higher spins tend to produce higher frequencies (e.g. see Appendix D, Table II of \cite{LISA_spectroscopy}); this is not true in general for counter-rotating modes, which were however ignored in this study given that they are not expected to strongly contribute for most of astrophysical events. As a consequence most of the biases on the spin will tend to be positive, which indeed is the case (82\% in our population).
A combination of biased $M_f$ and $a_f$ produces the \emph{GR} signal that best reproduces the injected \emph{nGR} waveform. 
The relevant figures are reported in Table~\ref{tab:stealth}, showing that $57\%$ of the total events show deviations in the final mass larger than $3\sigma$. Of this subset of events, $29\%$ ($76\%$) have SNR $\leq 30$ ($\leq 50$).
Thus for most of these signals one could check for consistency between the estimates of the remnant properties using separately the inspiral and the merger-ringdown parts of the signals. This is what is done in the standard IMR consistency test employed by the LVC~\cite{IMR_consistency_test1, IMR_consistency_test2}, which would be able to flag a discrepancy with GR predictions.

\begin{table}
\caption{Fraction of signals -- out of 60 \emph{nGR} injections recovered using a \emph{GR} template -- showing an effect size of the absolute values of the parameters $\Delta M_f$ and $\Delta a_f$ (first column), defined through Eq.~(\ref{eqn:effect_size_bias}), greater than 1 (second column), 2 (third column), and 3 (fourth column) .}
\begin{indented}
\item[] \begin{tabular}{@{}cccc}
\hline 
  & $\geq 1$  & $\geq 2$ & $\geq 3$ \\
\cline{1-4}
$|\Delta M_f|$  &  88\%   &   70\%   &   57\%    \\
$|\Delta a_f|$  &  12\%   &   3\%    &   0\%     \\
\end{tabular}
\end{indented}
\label{tab:stealth}
\end{table}

\section{Summary}\label{sec:Summary_section}

We investigated a phenomenological model of QBH based on the area quantisation conjecture, which in the heuristic formulation presented in~\cite{FOIT-KLEBAN} assigns a QNM spectrum with an explicit dependence on the quantum parameter $\alpha$. 
Employing a full Bayesian time-domain data analysis framework based on the \emph{pyRing} infrastructure \cite{CARULLO-DELPOZZO-VEITCH, Isi:2019aib, O3a-TGR-PAPER}, we showed how information from current events detected by LIGO and Virgo does not allow to discriminate between this model and the predictions of GR for the ringdown of a BBH coalescence. 
We demonstrated how taking into account the full correlation structure of the parameter space significantly impacts on the inference of such an alternative model. 
We gave an explicit example of the pitfalls that one may incur if only the one-dimensional distribution of 
$\alpha$ is considered as a figure of merit, as opposed to the usage of Bayes factors, that instead are capable of correctly discriminate between competing models and identify which one is best supported by the data.
Performing an injection study at the LIGO and Virgo design sensitivity, we demonstrated that $\mathcal{O}(20)$ detections of GW150914-like events can confidently discriminate among GR and the quantum-area proposal, providing a measurement of $\alpha$ in case such model is preferred. 
We also quantified the stealth biases that arise in standard parameter estimation if this model is correct, but not taken into account.
Such biases can be significant already for individual loud events, thus a violation of the standard classical GR emission would be flagged already from IMR consistency tests  routinely performed by the LVC~\cite{IMR_consistency_test1, IMR_consistency_test2, TGR-LVC2016}. 
Future detectors~\cite{ET, CE, LISA} could boost the constraints on $\alpha$ obtainable from a single event, hence decreasing the number of observations required to claim a confident detection of non-GR effects such as the quantisation of the horison area.
Nonetheless, we stress that to conclusively exclude any other possible systematic effects and to corroborate the evidence of GR violation one should look at the cumulative Bayes factor derived from a population of events.

The near-horizon space-time dynamics seems at the moment one of the only avenues to probe quantum effects of gravity. 
We presented a robust and effective methodology aimed at detecting and inferring such and any other GR-violating physical effects from the observation of ringdown signals from GWs, in the hope that it will spark interest in the development of ringdown models in alternative theories of gravity. 
In fact, the model that we concentrated on and tested against observational data relies on several assumptions~\cite{FOIT-KLEBAN}. Among them, those of particular relevance are: i) the light ring structure is affected by the area quantisation conjecture, reflecting on the frequency of the dominant mode, thus offering a way to test this model; ii) the ringdown emission is a coherent superposition of single-graviton transitions; iii) finally, the area quantisation conjecture is assumed to be valid.
Nevertheless, the framework presented here can straightforwardly include more involved models, such as the one presented in~\cite{CARDOSO-FOIT-KLEBAN} and \cite{AGULLO2020}. 
Future possible extensions of this work may include a reanalysis of the LIGO-Virgo events by assuming specific values of $\alpha$ proposed in the literature, a completion of the FK heuristic model by deriving a prediction for $\tau_0$ within the quantum-area proposal, and an implementation of the more realistic quantum ringdown model presented in~\cite{CARDOSO-FOIT-KLEBAN}, which extends the model presented in~\cite{FOIT-KLEBAN} considering the additional feature of ringdown echoes generated by the lack of a perfect-absorbing horizon.
\\

\textit{Acknowledgments --}
The authors would like to thank Giancarlo Cella for useful discussions and Adrian Ka Wai Chung for comments on the manuscript. This work benefited from discussions within the \textit{strong-field} working group of the LIGO/Virgo collaboration.
JV was partially supported by STFC grant ST/K005014/2. 
This research has made use of data, software and/or web tools obtained from the Gravitational Wave Open Science Center (https://www.gw-openscience.org), a service of LIGO Laboratory, the LIGO Scientific Collaboration and the Virgo Collaboration. LIGO is funded by the U.S. National Science Foundation. Virgo is funded by the French Centre National de Recherche Scientifique (CNRS), the Italian Istituto Nazionale di Fisica Nucleare (INFN) and the Dutch Nikhef, with contributions by Polish and Hungarian institutes.\\

\textit{Data availability statement --}
The data that support the findings of this study are available upon reasonable request from the authors.\\

\textit{References}
\bibliographystyle{iopart-num}
\bibliography{AQ_References}
\end{document}